\documentclass[journal]{IEEEtran}
\usepackage{amssymb}
\usepackage{amsbsy}
\usepackage{amsmath}
\usepackage{amsthm} 
\usepackage{setspace}
\usepackage{epsfig}
\usepackage{cite}
\usepackage{graphics}
\usepackage{epstopdf}
\usepackage{mathrsfs}
\usepackage[above]{placeins}
\usepackage{algorithm}
\usepackage{algpseudocode}
\usepackage{multirow}
\usepackage{algorithmicx}
\usepackage{tikz}

\begin{document}
\title{\huge{Deterministic Sampling Decoding: Where Sphere Decoding Meets Lattice Gaussian Distribution}}

\author{Zheng~Wang,~\IEEEmembership{Member, IEEE,}
        Cong~Ling,~\IEEEmembership{Member, IEEE}
        and Shi~Jin,~\IEEEmembership{Senior Member, IEEE}
\thanks{This work was supported in part by the open research fund of Key Laboratory of Dynamic Cognitive System of Electromagnetic Spectrum Space (Nanjing Univ. Aeronaut. Astronaut.), Ministry of Industry and Information Technology, Nanjing, 211106, China (No. KF20181913), the open research fund of National Mobile Communications Research Laboratory, Southeast University (No. 2019D04), the National Natural Science Foundation of China under Grant 61801216, in part by the Natural Science Foundation of Jiangsu Province under Grant BK20180420.


Z. Wang is with College of Electronic and Information Engineering, Nanjing University of Aeronautics and Astronautics (NUAA), Nanjing, China; C. Ling is with the Department of Electrical and Electronic Engineering, Imperial
College London, London, SW7 2AZ, United Kingdom; S. Jin is with the National Mobile Communications
Research Laboratory, Southeast University, Nanjing 210096, China (e-mail: wznuaa@gmail.com, cling@ieee.org, jinshi@seu.edu.cn).


}}

\maketitle

\begin{abstract}
In this paper, the paradigm of sphere decoding (SD) based on lattice Gaussian distribution is studied, where the sphere radius $D>0$ in the sense of Euclidean distance is characterized by the initial pruning size $K>1$, the standard deviation $\sigma>0$ and a regularization term $\rho_{\sigma,\mathbf{y}}(\Lambda)>0$ ($\Lambda$ denotes the lattice, $\mathbf{y}$ is the query point). In this way, extra freedom is obtained for analytical diagnosis of both the decoding performance and complexity.
Based on it, the equivalent SD (ESD) algorithm is firstly proposed, and we show it is exactly the same with the classic Fincke-Pohst SD but characterizes the sphere radius with $D=\sigma\sqrt{2\ln K}$.
By fixing $\sigma$ properly, we show that the complexity of ESD measured by the number of visited nodes is upper bounded by $|S|<nK$, thus resulting in a tractable decoding trade-off solely determined by $K$.
In order to further exploit the decoding potential, the regularized SD (RSD) algorithm based on Klein's sampling probability is proposed, which achieves a better decoding trade-off than the equivalent SD by fully utilizing the regularization terms.
Moreover, besides the designed criterion of pruning threshold, another decoding criterion named as candidate protection is proposed to solve the decoding problems in the cases of small $K$, which generalizes both the regularized SD and equivalent SD from maximum likelihood (ML) decoding to bounded distance decoding (BDD). Finally, simulation results based on MIMO detection are presented to confirm the tractable decoding trade-off of the proposed lattice Gaussian distribution-based SD algorithms.
\end{abstract}

\IEEEpeerreviewmaketitle


\IEEEpeerreviewmaketitle

\section{Introduction}
\IEEEPARstart{A}{s} one of the core problems of lattice decoding, the closest vector problem (CVP) has wide applications in number theory, cryptography, and communications.
However, the dramatically increased system size also places a pressing challenge upon solving the CVP.
On one hand, the conventional decoding schemes like lattice-reduction-aided decoding show a substantial performance loss with the increment of system dimension \cite{Babai,Yao2002,Taherzadeh2007,JaldenJDMT,CongProxity}.
On the other hand, a number of maximum-likelihood (ML) decoding schemes that aim to reduce the computational complexity of sphere decoding (SD) turn out to be impractical due to the unaffordable complexity in large-scale systems \cite{Kannan,Agrell2002,HassibiExpected,DamenDetectionSearch}. As for those near-ML decoding schemes like fixed-complexity sphere decoding (FCSD), K-best decoder, etc., they are also inapplicable
due to the intensive complexity increment and terrible performance deterioration \cite{Hassibisoft,JaldenFSD,4154776,EmbeddingLuzzi}.
In this condition, a number of advanced decoding schemes have been proposed to either improve the performance or lower the complexity \cite{TabuSrinidhi,DaiL1,GaoX1,MSMIMO1,6008619}. Among them, sampling decoding has become a promising one, which performs lattice decoding by sampling from a discrete multidimensional Gaussian distribution \cite{HassibiMCMCnew,McmcDatta,XiaodongWangMultilevel,ChoiMCMC1}.

Typically, sampling decoding converts the conventional decoding problem into a sampling problem, where the optimal decoding solution with the smallest Euclidean distance naturally entails the largest probability to be sampled.
However, in sharp contrast with continuous Gaussian density, it is by no means trivial even for sampling from a low-dimensional discrete Gaussian distribution, which means sampling decoding chiefly lies on how to successfully sample over the target lattice Gaussian distribution. For this reason, the pioneer works of sampling decoding only perform the sampling over a discrete Gaussian-like distribution \cite{Klein,CongRandom,DerandomizedJ}.
On the other hand, the classic Markov chain Monte Carlo (MCMC) methods were introduced to perform the exact sampling though the mixing of the Markov chains \cite{ChenMCMC,ZhengWangMCMCLatticeGaussian,MCMCHaidongZhu}. Moreover, in \cite{ZhengWangTIT15}, the independent Metropolis-Hastings-Klein (IMHK) sampling algorithm with accessible convergence rate was given and was further adopted to sampling decoding in \cite{ZhengWangTIT17}, thus leading to a tractable sampling decoding by adjusting the number of Markov moves.
Nevertheless, sampling decoding suffers from the inherent randomness during the sampling. On one hand, the possibility of missing the optimal decoding solution does always exist, rendering the inevitable performance loss. On the other hand, because of the independent and identically distributed (i.i.d.) samplings, lots of complexities in sampling are actually spent on those same calculations for many times, which means enormous complexity is wasted.

In this paper, to overcome the randomness during the sampling decoding, a deterministic sampling decoding scheme based on lattice Gaussian distribution is studied.
Specifically, since the optimal decoding solution has the largest sampling probability in the unimodal lattice Gaussian distribution, if lattice points with sampling probabilities larger than a certain level can be deterministically obtained, then the optimal decoding solution will be easily obtained.
To this end, deterministic sampling decoding can be viewed as sphere decoding, but the lattice point enumeration is performed according to the probability constraint by lattice Gaussian distribution rather than the traditional sphere constraint by Euclidean distance \cite{JaldenSphere,MuruganFrameworkSD,HassibiPruning,5773010}.
This is quite crucial for the study of CVP as extra freedom can be obtained by lattice Gaussian distribution to interpret the conventional sphere decoding, which provides a more comprehensive way for the analytical diagnosis of the decoding trade-off between performance and complexity.
Although Klein mentioned a deterministic scheme very briefly in \cite{Klein}, it does not seem to allow for an efficient implementation. Meanwhile, the heuristic implementation of deterministic sampling in \cite{DerandomizedJ} is hard to characterize the decoding trade-off in theory.
To summarize, we advance the state of the art of solving CVP in the following several fronts.

Firstly, we introduce the concept of lattice Gaussian distribution to CVP, where the related ML decoding becomes to find the lattice point with the largest sampling probability. Then, to realize the ML decoding, the sphere radius based on lattice Gaussian distribution is derived as $D_{\text{LGD}}=\sigma\sqrt{2\ln \frac{K}{\rho_{\sigma,\mathbf{y}}(\Lambda)}}$, where $\rho_{\sigma,\mathbf{y}}(\Lambda)$ is a Gaussian scalar conditioned on the lattice $\Lambda$ and the query point $\mathbf{y}$.
Interestingly, compared to the conventional sphere radius $D>0$, the sphere radius $D_{\text{LGD}}$ based on lattice Gaussian distribution is not only characterized by two new parameters named as initial pruning size $K>1$ and standard deviation $\sigma>0$, but also takes the Gaussian scalar $\rho_{\sigma,\mathbf{y}}(\Lambda)>0$ into account. By exploiting the relationship between the query point $\mathbf{y}$ and lattice $\Lambda$, the Gaussian scalar $\rho_{\sigma,\mathbf{y}}(\Lambda)$ actually serves as a regularization term to adjust the sphere radius $D_{\text{LGD}}$.
Unfortunately, since $\rho_{\sigma,\mathbf{y}}(\Lambda)$ is hard to compute or factorize, how to design a sphere decoding algorithm with sphere radius $D_{\text{LGD}}$ turns out to be quite challenging.

Secondly, following the clue of sphere radius based on lattice Gaussian distribution, the equivalent sphere decoding (ESD) algorithm is proposed based on a designed criterion named as \emph{pruning threshold}. Moreover, we demonstrate that it is exactly the same with the classic Fincke-Pohst SD with sphere radius $D=\sigma\sqrt{2\ln K}\triangleq D_{\text{equivalent}}$.
According to it, great flexibility is achieved as one can simply fix $\sigma$ and enjoy the decoding trade-off through tuning $K$. More importantly, the potential from the parameter $\sigma$ could be further exploited, and we show that by letting $\sigma=\min_{i}|r_{i,i}|/(2\sqrt{\pi})$, the number of visited nodes will be upper bounded by $|S|< nK$, which corresponds to the sphere radius $D_{\text{equivalent}}=\sqrt{\ln K/2\pi}\min_{i}|r_{i,i}|$ ($r_{i,i}$ is the $i$th diagonal element of the upper triangular matrix $\mathbf{R}$ with  $\mathbf{B}=\mathbf{QR}$).
Note that given the lattice basis $\mathbf{B}$, such an exact decoding trade-off between performance and complexity is exclusively determined by $K$.
Compared to sphere radius $D_{\text{LGD}}$, the equivalent SD (i.e., Fincke-Pohst SD) is essentially a reduced version without considering the impact of the given query point $\mathbf{y}$. In other words, equivalent SD only amounts to the special case with the Gaussian scalar $\rho_{\sigma,\mathbf{y}}(\Lambda)=1$.

Thirdly, in order to explore the decoding potential from the regularization term, the regularized sphere decoding (RSD) algorithm is proposed, which approximates the lattice Gaussian distribution through Klein's sampling probability.
In particular, by recursively performing the decoding layer by layer through the designed pruning threshold, the proposed regularized SD based on Klein's sampling probability is able to obtain the lattice points within sphere radius $D_{\text{regularized}}$. Meanwhile, its decoding gain by means of sphere radius over the equivalent SD (i.e., $G\triangleq\frac{D_{\text{regularized}}}{D_{\text{equivalent}}}$) is also derived.
Besides, with $\sigma=\min_{i}|r_{i,i}|/(2\sqrt{\pi})$, we show that the complexity of the regularized SD is still upper bounded by $|S|_{\text{regularized}}< nK$, which therefore leads to a better decoding trade-off than the equivalent SD.
Furthermore, as for solving CVP, we also demonstrate that the required initial pruning size $K$ of the regularized SD is upper bounded by $e^{2\pi d^2(\mathbf{\Lambda},\mathbf{y})/\min^2_ir_{i,i}}$, which corresponds to the complexity upper bounded by $|S|<n\cdot e^{2\pi d^2(\mathbf{\Lambda},\mathbf{y})/\min^2_ir_{i,i}}$ ($d(\mathbf{\Lambda},\mathbf{y})$ represents the Euclidean distance between the query point $\mathbf{y}$ and lattice $\Lambda$).

Finally, we try to solve a latent issue for both equivalent SD and regularized SD as no eligible lattice points will be outputted if the sphere radius $D_{\text{equivalent}}$ or $D_{\text{regularized}}$ is set smaller than $d(\mathbf{\Lambda},\mathbf{y})$. Similarly, this problem also exists in Fincke-Pohst SD so that suboptimal decoding scheme is preferred to offer an initial sphere radius. To this end, another decoding criterion named as \emph{candidate protection} is proposed to output a number of alternative decoding solutions based on the saved candidate nodes.
Note that the criterion of candidate protection is well compatible with the pruning threshold, and the decoding complexity including candidate protection in the regularized SD is still bounded by $|S|_{\text{regularized}}< nK$. We point out that candidate protection works in both equivalent SD and regularized SD but here we only consider the case of regularized SD for short.
By doing this, we effectively transfer the equivalent SD (i.e., Fincke-Pohst SD) and regularized SD from ML decoding into bounded distance decoding (BDD). In fact, with $K=1$, the decoding performance of Babai's nearest plane algorithm (also known as successive interference cancelation (SIC) in MIMO detection \cite{Babai}) will be achieved by the regularized SD with complexity $|S|=n$.
Moreover, we emphasize that such a change is rather crucial for the development of SD: as an ML decoding scheme, it has been ignored for a long time due to the rise of decoding large-scale problems (i.e., massive multiple-input multiple output (MIMO) systems) \cite{5595728,6736761,6375940,6457363}.

The rest of this paper is organized as follows. Section II introduces the lattice Gaussian distribution and briefly reviews the basics of lattice decoding based on SD.
In Section III, the concept of lattice Gaussian distribution is introduced to CVP for the first time, where the related sphere radius based on lattice Gaussian distribution is derived. From it, the equivalent SD algorithm is proposed, followed by the related analysis in both decoding performance and complexity.
In Section IV, the proposed regularized SD based on Klein's sampling probability is presented to achieve a better decoding trade-off than the equivalent SD. In Section V, another decoding criterion known as candidate protection is given to address the decoding problems of both equivalent SD and regularized SD in the cases of a small $K$, thus establishing a whole framework of SD. Simulation results for MIMO detection are shown in Section VI. Finally, Section VII concludes the paper.

\emph{Notation:} Matrices and column vectors are denoted by upper
and lowercase boldface letters, and the transpose, inverse, pseudoinverse
of a matrix $\mathbf{B}$ by $\mathbf{B}^T, \mathbf{B}^{-1},$ and
$\mathbf{B}^{\dag}$, respectively. We use $\mathbf{b}_i$ for the $i$th
column of the matrix $\mathbf{B}$, $b_{i,j}$ for the entry in the $i$th row
and $j$th column of the matrix $\mathbf{B}$. $\lceil x \rfloor$ denotes rounding to
the integer closest to $x$. If $x$ is a complex number, $\lceil x \rfloor$
rounds the real and imaginary parts separately.
Finally, in this paper, the complexity of SD is evaluated by the number of visited nodes (i.e., $|S|$) during the decoding along the tree traversal.
Meanwhile, the computational complexity is measured by the number of arithmetic operations (additions, multiplications, comparisons, etc.).

\newtheorem{my1}{Lemma}
\newtheorem{my2}{Theorem}
\newtheorem{my3}{Definition}
\newtheorem{my4}{Proposition}
\newtheorem{my5}{Remark}
\newtheorem{my6}{Conjection}
\newtheorem{my7}{Corollary}

\section{Preliminaries}
In this section, we introduce the background and mathematical tools needed to describe and analyze the proposed SD based on lattice Gaussian distribution.

\subsection{Lattice Decoding \& Fincke-Pohst Sphere Decoding}
Given the full $n\times n$ column-rank matrix $\mathbf{B}\in\mathbb{R}^{n\times n}$, the $n$-dimensional lattice $\Lambda$ generated by it is defined by
\begin{equation}
\Lambda=\{\mathbf{Bx}: \mathbf{x}\in \mathbb{Z}^n\},
\end{equation}
where $\mathbf{B}$ is called the lattice basis.
Consider the decoding of an $n \times n$ real-valued system. The extension to the complex-valued system is straightforward \cite{CongRandom,Xia1}. Let $\mathbf{x}\in \mathbb{Z}^n$ denote the transmitted signal.
The corresponding received signal $\mathbf{c}$ is given by
\begin{equation}
\mathbf{c}=\mathbf{B}\mathbf{x}+\mathbf{w}
\label{eqn:System Model}
\end{equation}
where $\mathbf{w}$ is the noise vector with zero mean and variance $\sigma_w^{2}$.
Typically, the conventional maximum likelihood (ML) reads
\begin{equation}
\widehat{\mathbf{x}}=\underset{\mathbf{x}\in \mathbb{Z}^{n}}{\operatorname{arg~min}} \, \|\mathbf{B}\mathbf{x}-\mathbf{c}\|^2
\label{eqn:ML Decoding}
\end{equation}
where $\| \cdot \|$ denotes the Euclidean norm. Clearly, the ML decoding in above MIMO systems corresponds to the CVP in lattice \cite{DamenDetectionSearch}.


\begin{figure}[t]
\vspace{-2em}
\begin{center}
\hspace{-1em}\includegraphics[width=3.6in,height=2.4in]{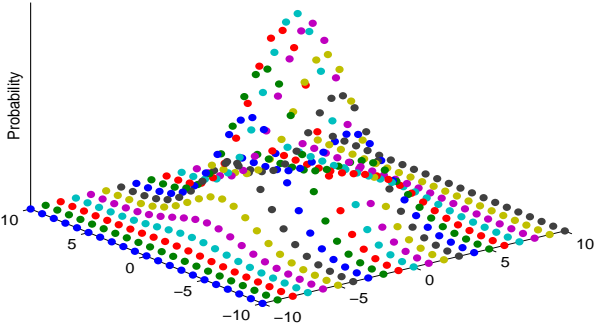}
\end{center}
\vspace{-2em}
  \caption{Illustration of a two-dimensional lattice Gaussian distribution.}
  \label{simulation x}
\end{figure}

Here, for notational simplicity, QR-decomposition with $\mathbf{B}=\mathbf{QR}$ is applied and we express the system model in (\ref{eqn:System Model}) as
\begin{equation}
\mathbf{y}=\mathbf{Q}^T\mathbf{c}=\mathbf{R}\mathbf{x}+\mathbf{n},
\label{eqn:QR}
\end{equation}
where $\mathbf{Q}$ is an orthogonal matrix and $\mathbf{R}$ is an upper triangular matrix. Accordingly, the ML decoding in (\ref{eqn:ML Decoding}) becomes
\begin{equation}
\widehat{\mathbf{x}}_{\text{ML}}=\underset{\mathbf{x}\in \mathbb{Z}^{n}}{\operatorname{arg~min}} \, \|\mathbf{R}\mathbf{x}-\mathbf{y}\|^2.
\label{eqn:ML DecodingQR}
\end{equation}


In the classic Babai's nearest plane algorithm, $\widehat{x}_i$ is decoded in a backwards order layer by layer (i.e., $i=n,n-1,\ldots,1$) by direct rounding
\begin{equation}
\widehat{x}_i=\lceil\widetilde{x}_i\rfloor,
\label{rounding}
\end{equation}
where
\begin{equation}
\widetilde{x}_i=\frac{y_i-\sum^n_{j=i+1}r_{i,j}\widehat{x}_j}{r_{i,i}}.
\label{eqn:noise effection in determination}
\end{equation}

On the other hand, to achieve the ML decoding performance, the classic Fincke-Pohst sphere decoding (SD) was proposed to enumerate all the possible lattice points within a sphere radius $D>0$ \cite{FPSD}
\begin{equation}
\|\mathbf{Rx}-\mathbf{y}\|\leq D.
\label{sd1}
\end{equation}
Specifically, given $\widetilde{x}_i$, the searching space of $\widehat{x}_i$ in the recursive decoding from layer $n$ to $1$ can be written as
\begin{equation}
|\widehat{x}_i-\widetilde{x}_i|_{\text{Fincke-Pohst}}\hspace{-.2em}\leq \hspace{-.2em}\sqrt{D^2\hspace{-.2em}-\hspace{-.2em}\sum_{j=i+1}^n \left| y_j\hspace{-.2em}-\hspace{-.2em}\sum_{l=j}^n r_{j,l}\widehat{x}_l \right|^2}/|r_{i,i}|,
\label{sdsame}
\end{equation}
where candidate node $\widehat{x}^j_i$ satisfying (\ref{sdsame}) will be saved. Here, $\widehat{x}^j_i$ denotes the $j$th closest integer candidate node to $\widetilde{x}_i$.

Finally, among the collected lattice points, the one with the closest Euclidean distance $\|\mathbf{R\widehat{x}}-\mathbf{y}\|$ will be outputted as the decoding solution.
Unfortunately, SD entails an exponentially increased complexity with the sphere radius $D$, thereby making it unaffordable especially in high-dimensional systems \cite{Phost}.
Additionally, the sphere radius $D$ should be selected carefully since a large one would lead to considerable complexity waste while no eligible lattice points would be yielded with a small choice of $D$.

\subsection{Lattice Gaussian Distribution \& Klein's Sampling Probability}
Given the lattice $\Lambda=\{\mathbf{Rx}: \mathbf{x}\in \mathbb{Z}^n\}$, define the Gaussian function centered at $\mathbf{y}\in \mathbb{R}^n$ for standard deviation $\sigma>0$ as
\begin{equation}
\rho_{\sigma, \mathbf{y}}(\mathbf{z})=e^{-\frac{\|\mathbf{z}-\mathbf{y}\|^2}{2\sigma^2}},
\end{equation}
for all $\mathbf{z}\in\mathbb{R}^n$. When $\mathbf{y}$ or $\sigma$ are not specified, it is assumed that they are $\mathbf{0}$ and $1$ respectively. Then, the \emph{discrete Gaussian distribution} over $\Lambda$ is defined as \cite{MicciancioGaussian}
\begin{equation}
D_{\Lambda,\sigma,\mathbf{y}}(\mathbf{x})=\frac{\rho_{\sigma, \mathbf{y}}(\mathbf{Rx})}{\rho_{\sigma, \mathbf{y}}(\Lambda)}=\frac{e^{-\frac{1}{2\sigma^2}\parallel \mathbf{Rx}-\mathbf{y} \parallel^2}}{\sum_{\mathbf{x} \in \mathbb{Z}^n}e^{-\frac{1}{2\sigma^2}\parallel \mathbf{Rx}-\mathbf{y} \parallel^2}}
\label{lattice gaussian distribution}
\end{equation}
for all $\mathbf{x}\in \mathbb{Z}^n$, where $\rho_{\sigma, \mathbf{y}}(\Lambda)\triangleq \sum_{\mathbf{\mathbf{Rx}}\in\Lambda}\rho_{\sigma, \mathbf{y}}(\mathbf{Rx})$ is a Gaussian scalar to ensure a probability distribution.

Due to the central role of the lattice Gaussian distribution playing in various research fields, sampling from lattice Gaussian distribution (known as lattice Gaussian sampling) becomes an important but challenging problem. As an approximation for lattice Gaussian sampling, Klein's sampling algorithm was proposed in \cite{Klein}, which is able to sample from a discrete multi-dimensional Gaussian-like distribution.
The mechanism of Klein's sampling algorithm can be viewed as a statistic variant of Babai's nearest plane algorithm.
Specifically, $\widehat{x}_i$ is randomly chosen from the following 1-dimensional conditional lattice Gaussian distribution
{\allowdisplaybreaks\begin{flalign}
\widehat{x}_i\sim p(\widehat{x}_i)&\triangleq D_{\mathbb{Z},\sigma_i,\widetilde{x}_i}(x_i=\widehat{x}_i)\notag\\
&=\frac{e^{-\frac{1}{2\sigma^2_i}\|\widehat{x}_i-\widetilde{x}_i\|^2}}{\sum_{\widetilde{x}_i\in\mathbb{Z}}e^{-\frac{1}{2\sigma^2_i}\|\widehat{x}_i-\widetilde{x}_i\|^2}}
\label{piq19}
\end{flalign}}in a backwards order with $\sigma_i=\frac{\sigma}{|r_{i,i}|}$, which makes the sample $\widehat{\mathbf{x}}$ obey the following Klein's sampling probability
{\allowdisplaybreaks\begin{flalign}
P_{\text{Klein}}(\widehat{\mathbf{x}})&=\prod^n_{i=1}D_{\mathbb{Z},\sigma_{n-i+1},\widetilde{x}_{n-i+1}}(x_{n-i+1})\notag\\
&=\frac{e^{-\frac{1}{2\sigma^2}\parallel \mathbf{R\widehat{x}}-\mathbf{y} \parallel^2}}{\prod^n_{i=1}\sum_{\widetilde{x}_{n-i+1}\in\mathbb{Z}}e^{-\frac{1}{2\sigma^2_{n-i+1}}\|x_{n-i+1}-\widetilde{x}_{n-i+1}\|^2}}\notag\\
&=\frac{\rho_{\sigma, \mathbf{y}}(\mathbf{R\widehat{x}})}{\prod^n_{i=1}\rho_{\sigma_{n-i+1}, \widetilde{x}_{n-i+1}}(\mathbb{Z})}.
\label{klein distribution}
\end{flalign}}

By sampling from Klein's sampling probability, Klein's sampling decoding and randomized sampling decoding schemes were proposed. However, it has been demonstrated in \cite{Trapdoor} that $P_{\text{Klein}}(\mathbf{x})$ can be close to $D_{\Lambda,\sigma,\mathbf{y}}(\mathbf{x})$ only when $\sigma$ is sufficiently large. Unfortunately, such a condition is extremely stringent, rendering it inapplicable in many cases of interest.

\renewcommand{\algorithmicrequire}{\textbf{Input:}}  
\renewcommand{\algorithmicensure}{\textbf{Output:}} 

\begin{algorithm}[t]
\caption{Klein's Algorithm}
\begin{algorithmic}[1]
\Require
$\mathbf{B}, \sigma, \mathbf{c}$
\Ensure
$\mathbf{Bx}\in\Lambda$
\State let $\mathbf{B}=\mathbf{QR}$ and $\mathbf{y}=\mathbf{Q}^{\dag}\mathbf{c}$
\For {$i=n$,\ \ldots,\ 1}
\State let $\sigma_i=\frac{\sigma}{|r_{i,i}|}$ and $\widetilde{x}_i=\frac{y_i-\sum^n_{j=i+1}r_{i,j}x_j}{r_{i,i}}$
\State sample $x_i$ from $D_{\mathbb{Z},\sigma_i,\widetilde{x}_i}$
\EndFor
\State return $\mathbf{Bx}$
\end{algorithmic}
\end{algorithm}

\section{Sphere Decoding Revisited Based on Lattice Gaussian Distribution}
In this section, inspired by SD based on lattice Gaussian distribution, the classic Fincke-Pohst SD is restudied in a novel way. The equivalent SD algorithm is proposed, which is exactly the same with Fincke-Pohst SD but parameterized by the initial pruning size $K$ and standard deviation $\sigma$.
Based on it, both decoding performance and complexity turn out to be analytical, thus leading to a tractable trade-off.

\subsection{SD based on Lattice Gaussian Distribution}
From (\ref{lattice gaussian distribution}), since lattice Gaussian distribution is centered at the query point $\mathbf{y}$, the closest lattice point $\mathbf{Rx}_{\text{ML}}$ to $\mathbf{y}$ is naturally assigned the largest sampling probability. Therefore, by multiple i.i.d. samplings, the CVP in (\ref{eqn:ML Decoding}) can be solved as $\mathbf{x}_{\text{ML}}$ is most likely to be returned.
More specifically, from the point of view of lattice Gaussian distribution, the the conventional ML decoding in (\ref{eqn:ML Decoding}) becomes
\begin{equation}
\widehat{\mathbf{x}}_{\text{ML}}=\underset{\mathbf{x}\in \mathbb{Z}^{n}}{\operatorname{arg~max}} \, D_{\Lambda,\sigma,\mathbf{y}}(\mathbf{x}),
\label{eqn:ML DecodingQR1}
\end{equation}
which converts the decoding problem into a sampling problem. It has been demonstrated that lattice Gaussian sampling is equivalent to CVP via a polynomial-time dimension-preserving reduction \cite{DGStoCVPSVP}.

Compared to those existing decoding solutions by Euclidean distance, decoding by sampling has promising advantages.
Firstly, the standard deviation $\sigma$ can be optimized to improve the sampling probability of the target point, resulting in a better decoding performance. Secondly, by adjusting the number of sampling times, the sampler decoding enjoys a flexible trade-off between performance and complexity.
Nevertheless, the problem of sampler decoding chiefly lies on how to perform the sampling over the target lattice Gaussian distribution. To this end, the classic MCMC method is introduced, where lattice Gaussian sampling can be achieved through the Markov mixing \cite{ZhengWangTIT15,ZhengWangTIT17}. However, due to the inherent randomness during the sampling, considerable performance loss and complexity waste are inevitable in sampling decoding.

To overcome the randomness for a better decoding trade-off, it is feasible to deterministically collect all the lattice points whose sampling probabilities satisfy
\begin{equation}
D_{\Lambda,\sigma,\mathbf{y}}(\mathbf{x})=\frac{e^{-\frac{1}{2\sigma^2}\parallel \mathbf{Rx}-\mathbf{y} \parallel^2}}{\sum_{\mathbf{x} \in \mathbb{Z}^n}e^{-\frac{1}{2\sigma^2}\parallel \mathbf{Rx}-\mathbf{y} \parallel^2}}\geq\frac{1}{K}
\label{lattice gaussian distribution1a}
\end{equation}
with the \emph{initial pruning size} $K>1$. In this way, we only need to pay attention on those lattice points with sampling probabilities above a certain level, i.e., $1/K$.
Because the lattice Gaussian distribution $D_{\Lambda,\sigma,\mathbf{y}}(\mathbf{x})$ is unimodal, the closest lattice point will be outputted by simply letting $D_{\Lambda,\sigma,\mathbf{y}}(\mathbf{x}_{\text{ML}})\geq1/K$.
Intuitively, such a deterministic sampling decoding based on (\ref{lattice gaussian distribution1a}) corresponds to enumerating all the lattice points within the sphere radius $D_{\text{LGD}}$
\begin{equation}
\|\mathbf{Rx}-\mathbf{y}\|\leq \sigma\sqrt{2\ln \frac{K}{\rho_{\sigma,\mathbf{y}}(\Lambda)}}\triangleq D_{\text{LGD}},
\label{lattice gaussian distribution1c}
\end{equation}
which is analogous to the sphere radius of Fincke-Pohst SD in (\ref{sd1}).
Different from Fincke-Pohst SD, the Gaussian scalar $\rho_{\sigma,\mathbf{y}}(\Lambda)>0$ is introduced, which takes the given query point $\mathbf{y}$ into account.

Unfortunately, the Gaussian scalar $\rho_{\sigma,\mathbf{y}}(\Lambda)>0$ is difficult to compute and factorize. In this condition, how to design the decoding algorithm with sphere radius $D_{\text{LGD}}$ turns out to be quite challenging. Nevertheless, the sphere radius $D_{\text{LGD}}$ in (\ref{lattice gaussian distribution1c}) still provides a meaningful clue to reconsider the sphere radius in SD, namely, what is the relationship between sphere radii $D$ and $D_{\text{LGD}}$? To answer this question, in what follows, we show that the classic Fincke-Pohst SD is actually a reduced version of SD based on lattice Gaussian distribution, where the Gaussian scalar $\rho_{\sigma,\mathbf{y}}(\Lambda)$ essentially serves as a regularization term for the sphere radius.

\subsection{Equivalent SD Algorithm}
Motivated by SD based on lattice Gaussian distribution, we now present the proposed equivalent SD algorithm. First of all, the tree-search decoding structure in Fincke-Pohst SD is retained and the decoding is still performed layer by layer in a backwards order.
Then, based on the initial pruning size $K>1$ set at the beginning, with respect to each integer candidate node $\widehat{x}^j_i$ in the tree traversal, its \emph{pruning size}
is defined as
\begin{equation}
K(\widehat{x}^j_i)\triangleq K(\underline{\widehat{x}^j_{i}})\cdot f(\widehat{x}^j_i)
\label{prunethresholdla}
\end{equation}
with
\begin{equation}
f(\widehat{x}^j_i)\triangleq e^{-\frac{1}{2\sigma^2_i}\|\widehat{x}^j_i-\widetilde{x}_i\|^2},
\label{fn1}
\end{equation}
and $\widehat{x}^j_i$ will be retained only if the following \emph{pruning threshold}
\begin{equation}
K(\widehat{x}^j_i)\geq 1
\label{prunethreshold1b}
\end{equation}
is satisfied. Otherwise, $\widehat{x}^j_i$ will be discarded and the decoding moves to the next layer based on those survived candidate nodes.
Here, $\underline{\widehat{x}^j_{i}}$ denotes the parent node of $\widehat{x}^j_{i}$ at the last decoding layer $i+1$.
Clearly, several children candidate nodes $\widehat{x}^{j}_{i}$ may have a same parent node $\underline{\widehat{x}^j_{i}}$ and $K(\underline{\widehat{x}^j_n})=K$.
From (\ref{prunethresholdla}) and (\ref{prunethreshold1b}), the pruning threshold can be further expressed by
\begin{equation}
f(\widehat{x}^j_i)\geq\frac{1}{K(\underline{\widehat{x}^j_i})}=\frac{1}{K\cdot f(\widehat{x}^j_{i+1})\cdots f(\widehat{x}^j_{n})},
\label{b12b}
\end{equation}
where the initial pruning size $K$ essentially serves as a parameter to adjust the pruning threshold. Intuitively, a larger size $K$ corresponds to a smaller pruning threshold at each layer, thereby saving more decoding candidates
at the end.
Finally, the lattice point with the closest Euclidean distance among the candidate list will be outputted as the decoding solution.
To summarize, the operation of equivalent SD is presented in Algorithm 2.

\begin{my1}
Given the initial pruning size $K>1$, lattice points within sphere radius
\end{my1}
\vspace{-1em}
\begin{equation}
\|\mathbf{Rx}-\mathbf{y}\|\leq \sigma\sqrt{2\ln K}
\label{b12a}
\end{equation}
\emph{will be obtained by the proposed equivalent SD.}
\begin{proof}
To start with, (\ref{b12a}) can be rewritten as
\begin{equation}
e^{-\frac{1}{2\sigma^2}\|\mathbf{Rx}-\mathbf{y}\|^2}\geq \frac{1}{K},
\label{b12c}
\end{equation}
which can be further expressed by factorization as
\begin{equation}
\prod^n_{i=1}e^{-\frac{1}{2\sigma^2_{n-i+1}}\|\widehat{x}_{n-i+1}-\widetilde{x}_{n-i+1}\|^2}\hspace{-.5em}=\prod^n_{i=1}f(\widehat{x}_{n-i+1})\geq \frac{1}{K}.
\label{b12d}
\end{equation}

Then, in order to collect lattice points satisfying (\ref{b12d}), considering the fact that $0<f(\cdot)\leq1$, $f(\widehat{x}_i)$ should fulfill the following requirement
{\allowdisplaybreaks\begin{flalign}
f(\widehat{x}_i)&\geq\frac{1}{K\cdot\prod_{j\neq i}f(\widehat{x}_j)}\notag\\
&\geq\frac{1}{K\cdot f(\widehat{x}_{i+1})\cdots f(\widehat{x}_{n})}
\end{flalign}}for $1\leq i\leq n$, which exactly corresponds to the proposed pruning threshold in (\ref{b12b}).
\end{proof}

Based on Lemma 1, we now verify the equivalence of Fincke-Pohst SD and the proposed equivalent SD by showing they have the same searching space of $\widehat{x}_i$ at each decoding layer.


\begin{my2}
The Fincke-Pohst SD is exactly the same with the proposed equivalent SD with sphere radius
\end{my2}
\vspace{-1em}
\begin{equation}
D=\sigma\sqrt{2\ln K}\triangleq D_{\text{equivalent}}.
\label{mp1}
\end{equation}
\begin{proof}
According to the pruning threshold in (\ref{b12b}), the searching space of $\widehat{x}_i$ given $\widetilde{x}_i$ can be derived as
{\allowdisplaybreaks\begin{flalign}
|\widehat{x}_i-\widetilde{x}_i|_{\text{equivalent}}&\leq \sqrt{2\sigma^2\ln K-\sum_{j=i+1}^n \left| y_j-\sum_{l=j}^n r_{j,l}\widehat{x}_l \right|^2}/|r_{i,i}|\notag\\
&=\sqrt{D^2-\sum_{j=i+1}^n \left| y_j-\sum_{l=j}^n r_{j,l}\widehat{x}_l \right|^2}/|r_{i,i}|,
\end{flalign}}which is exactly the boundary of $|\widehat{x}_i-\widetilde{x}_i|_{\text{Fincke-Pohst}}$ in (\ref{sdsame}) for $1\leq i\leq n$, completing the proof.
\end{proof}

\begin{algorithm}[t]
\caption{Equivalent SD Algorithm}
\begin{algorithmic}[1]
\Require
$K, \mathbf{R}, \mathbf{y}, \sigma=\min_{i}|r_{i,i}|/(2\sqrt{\pi}), L=\emptyset$
\Ensure
$\mathbf{Rx}\in\Lambda$
\State invoke \textbf{Function 1} with $i=n$ to decode layer by layer
\State add all the candidates $\widehat{\mathbf{x}}$'s generated by \textbf{Function 1} to $L$
\State output $\widehat{\mathbf{x}}=\underset{\mathbf{x}\in L}{\operatorname{arg~min}} \, \|\mathbf{y}-\mathbf{R}\mathbf{x}\|$ as the decoding solution
\end{algorithmic}
\end{algorithm}

\floatname{algorithm}{Function}

\newcounter{TempEqCnt}                         
\setcounter{TempEqCnt}{\value{algorithm}} 
\setcounter{algorithm}{0}

\begin{algorithm}[t]
\caption{Pruning Decoding at layer $i$ given $[\widehat{x}_n,\ldots,\widehat{x}_{i+1}]$}
\begin{algorithmic}[1]
\State compute $\widetilde{x}_i$ according to (\ref{eqn:noise effection in determination})
\State compute probability $f(\widehat{x}^j_i)$ by (\ref{fn1})
\State compute pruning size $K(\widehat{x}^j_i)$ according to (\ref{prunethresholdla})
\For {each specific integer candidate $\widehat{x}^j_i$}
\If {$K(\widehat{x}^j_i)<1$}
\State \hspace{-2em}prune $\widehat{x}^j_i$ from the tree-search decoding
\Else
\State {\hspace{-2em}save $\widehat{x}^j_i$} to form the decoding result $[\widehat{x}_n,\ldots,\widehat{x}_{i+1}, \widehat{x}^j_i]$
\If {$i=1$}
\State \hspace{-2em}output the candidate $\widehat{\mathbf{x}}$
\Else
\State \hspace{-2em}invoke \textbf{Function 1} to decode the next layer $i-1$
\EndIf
\EndIf
\EndFor
\end{algorithmic}
\end{algorithm}

\subsection{Trade-off Analysis Between Performance and Complexity}
Based on $D=\sigma\sqrt{2\ln K}$, extra freedom can be obtained in interpreting SD, which provides a feasible way for the analytical diagnose in both decoding performance and complexity.

\begin{my1}
In the equivalent SD, for each parent candidate node $\underline{\widehat{x}^j_{i}}$ with $K(\underline{\widehat{x}^j_i})\geq1$,
the number of its saved children candidate nodes at decoding layer $i$ satisfies
\end{my1}
\vspace{-1em}
\begin{equation}
K_{\text{save}}\leq K(\underline{\widehat{x}^j_i})
\label{pj1ab}
\end{equation}
if $\sigma<\min_i|r_{i,i}|/(2\sqrt{2\ln 2})$.
\begin{proof}
According to the pruning threshold given in (\ref{prunethreshold1b}), the condition shown in (\ref{pj1ab}) holds if and only if the $\lfloor K(\underline{\widehat{x}^j_i})+1\rfloor$th closest integer candidate to $\widetilde{x}_i$ will definitely be pruned, that is
\begin{equation}
K(\underline{\widehat{x}^j_i})f(\widehat{x}^{\lfloor K(\underline{\widehat{x}^j_i})+1 \rfloor}_i)< 1.
\label{wo1a}
\end{equation}

Then, because the distance $|\widehat{x}^j_i-\widetilde{x}_i|$ is bounded by
\begin{equation}
(j-1)\cdot\frac{1}{2}\leq|\widehat{x}^j_i-\widetilde{x}_i|\leq j\cdot\frac{1}{2},
\label{wo31}
\end{equation}
(\ref{wo1a}) can be achieved if
\begin{equation}
K(\underline{\widehat{x}^j_i})\cdot e^{-\frac{1}{8\sigma_i^2}(\lfloor K(\underline{\widehat{x}^j_i})+1\rfloor-1)^2}<1,
\label{wo4a}
\end{equation}
which corresponds to
\begin{equation}
\sigma^2<\frac{(\lfloor K(\underline{\widehat{x}^j_i})+1\rfloor-1)^2}{8\ln K(\underline{\widehat{x}^j_i})}\cdot r^2_{i,i}.
\label{wo5al}
\end{equation}

Moreover, it is easy to verify the lower bound of the right-hand term of (\ref{wo5al}) as
\begin{equation}
\frac{(\lfloor K(\underline{\widehat{x}^j_i})+1\rfloor-1)^2}{8\ln K(\underline{\widehat{x}^j_i})}\cdot r^2_{i,i}>\frac{1}{8\ln 2}\cdot r^2_{i,i},
\label{wo5ab}
\end{equation}
which means (\ref{wo5al}) is fulfilled if
\begin{equation}
\sigma<\min_i |r_{i,i}|/(2\sqrt{2\ln 2})
\end{equation}
for $1\leq i\leq n$.
\end{proof}

\begin{my1}
In the equivalent SD, for each parent candidate node $\underline{\widehat{x}^j_{i}}$ with $K(\underline{\widehat{x}^j_i})\geq1$,
the summation of pruning sizes of its children candidate nodes at decoding layer $i$ is non-increasing
\end{my1}
\vspace{-1em}
\begin{equation}
\sum_jK(\widehat{x}^j_i)< K(\underline{\widehat{x}^j_i})
\label{pj1a1}
\end{equation}
if $\sigma\leq\min_{i}|r_{i,i}|/(2\sqrt{\pi})$.
\begin{proof}
Based on the definition given in (\ref{prunethresholdla}), it follows that
{\allowdisplaybreaks\begin{flalign}
\sum_jK(\widehat{x}^j_i)&=K(\underline{\widehat{x}^j_i})\cdot\sum_jf(\widehat{x}^j_i)\notag\\
&<K(\underline{\widehat{x}^j_i})\cdot \sum_{\widehat{x}_i\in\mathbb{Z}}e^{-\frac{1}{2\sigma^2_i}\|\widehat{x}_i-\widetilde{x}_i\|^2}\notag\\
&\overset{(a)}{\leq} K(\underline{\widehat{x}^j_i})\cdot \sum_{\widehat{x}_i\in\mathbb{Z}}e^{-\frac{1}{2\sigma^2_i}\|\widehat{x}_i\|^2}\notag\\
&\overset{(b)}{=}K(\underline{\widehat{x}^j_i})\cdot \vartheta_3(|r_{i,i}|^2/2\pi\sigma^2)\notag\\
&\overset{(c)}{\approx}K(\underline{\widehat{x}^j_i}).
\label{pj1a1}
\end{flalign}}Here, inequality (a) recalls the following relationship
\begin{equation}
\rho_{\sigma,\mathbf{c}}(\Lambda)\leq\rho_{\sigma}(\Lambda),
\label{v1f}
\end{equation}
where the equality holds only when $\mathbf{c}\in\Lambda$. Inequality (b) invokes the \emph{Jacobi theta function} $\vartheta_3$ \cite{ConwayandSloane}
\begin{equation}
\vartheta_3(\nu)=\sum^{+\infty}_{n=-\infty}e^{-\pi\nu n^2},
\label{b16}
\end{equation}
and the approximation in (c) follows
\begin{equation}
\prod_{i=1}\vartheta_3(|r_{i,i}|^2/2\pi\sigma^2)\leq\vartheta_3(2)=1.0039\approx1
\label{bound7}
\end{equation}
for $\sigma\leq\min_{i}|r_{i,i}|/(2\sqrt{\pi})$ because $\vartheta_3(\nu)$ is monotone decreasing with $\nu>0$.
\end{proof}

Based on Lemmas 2 \& 3, the complexity of the equivalent SD can be derived by means of the number of visited nodes as follows.
\begin{my2}
In the equivalent SD, let $\sigma=\min_{i}|r_{i,i}|/(2\sqrt{\pi})$, the number of visited nodes denoted by $|S|$ is upper bounded by
\end{my2}
\vspace{-2em}
\begin{equation}
|S|< nK.
\end{equation}
\begin{proof}
To start with, according to Lemma 2, the number of saved candidate nodes at each decoding layer is upper bounded by the summation of pruning sizes at the previous layer, i.e.,
\begin{equation}
K^{\text{layer}\ i}_{\text{save}}=\sum K_{\text{save}}\leq \sum K(\underline{\widehat{x}^j_i})=K^{\text{layer}\ i+1}_{\text{pruning\ size}}.
\end{equation}
Meanwhile, from Lemma 3, because the summation of pruning sizes at each decoding layer is decreasing as
\begin{equation}
K^{\text{layer}\ 1}_{\text{pruning\ size}}<\ldots< K^{\text{layer}\ n}_{\text{pruning\ size}}< K^{\text{layer}\ n+1}_{\text{pruning\ size}}=K
\label{pj1pa}
\end{equation}
so that the number of visited nodes in the proposed equivalent SD is upper bounded by
\begin{equation}
|S|=\sum^n_{i=1}K^{\text{layer}\ i}_{\text{save}}\leq \sum^n_{i=1}K^{\text{layer}\ i+1}_{\text{pruning\ size}}< nK,
\end{equation}
completing the proof.
\end{proof}

The clear complexity upper bound for equivalent SD in Theorem 2 answers a fundamental question in the framework of sphere decoding since it is difficult to evaluate the incurred complexity based on the single parameter $D$.
In \cite{HassibiExpected}, an average version of $|S|$ for Fincke-Pohst SD was given, which was further improved through the analysis of its asymptotic behaviour in \cite{JaldenSphere}. However, they mainly focus on characterizing the mean and variance of the complexity for i.i.d. Gaussian lattice basis. In \cite{6006604}, the tail exponents of the SD complexity distribution was investigated for the complexity estimation. Nevertheless, the number of visited nodes $|S|$ in it is evaluated in a probabilistic form.
Other works about complexity of SD can be found in \cite{6216420,5429127}, which either takes the specific conditions from communications into account or considers the complexity in infinity-norm SD.

Subsequently, because the number of saved candidate nodes at decoding layer $i=1$ corresponds to the number of collected lattice points i.e., $K^{\text{layer}\ 1}_{\text{save}}=|L|$, we can easily arrive at the following result.
\begin{my7}
The number of lattice points collected by the proposed equivalent SD denoted by $|L|$ is upper bounded by
\end{my7}
\vspace{-1em}
\begin{equation}
|L|< K.
\end{equation}

From Theorem 2, with $\sigma=\min_{i}|r_{i,i}|/(2\sqrt{\pi})$, the complexity of the equivalent SD is upper bounded. This is rather important for the study of SD as the trade-off between decoding performance and complexity becomes analytical. To this end, one can simply fix $\sigma$ and enjoy the decoding trade-off between performance and complexity through the single tunable parameter $K$, which naturally
leads to the following Theorem.
\begin{my2}
In the proposed equivalent SD, let $\sigma=\min_{i}|r_{i,i}|/(2\sqrt{\pi})$, the tractable sphere radius $D_{\text{equivalent}}=\sqrt{\frac{\ln K}{2\pi}}\min_{i}|r_{i,i}|$ corresponds to the complexity upper bounded by $|S|< nK$.
\end{my2}

From Theorem 3, with the increase of sphere radius $D_{\text{equivalent}}$, the corresponding $K$ improves exponentially. Here, we emphasize the significance of LLL reduction, which is able to effectively improve $\min_{i}|r_{i,i}|$ (i.e., $\min_{i}\|\widehat{\mathbf{b}}_i\|$, $\mathbf{\widehat{b}}_i$'s are the Gram-Schmidt vectors of the lattice basis $\mathbf{B}$) through the matrix transformation \cite{LLLoriginal,CongProxity}.


\begin{my7}
With the help of LLL reduction (i.e., $\overline{\mathbf{R}}=\mathbf{RU}$), the proposed equivalent SD with $\sigma=\min_{i}|\overline{r}_{i,i}|/(2\sqrt{\pi})$ achieves a larger sphere radius due to
\end{my7}
\vspace{-1em}
\begin{equation}
\min_{i}|\overline{r}_{i,i}|\geq\min_{i}|r_{i,i}|.
\end{equation}

Subsequently, consider decoding the CVP problem with $D=d(\Lambda, \mathbf{y})$, the required initial pruning size $K$ as well as the complexity $|S|$ can be derived in the following.

\begin{my2}
The initial pruning size $K$ of solving the CVP by the proposed equivalent SD is $e^{\frac{2\pi d^2(\mathbf{\Lambda},\mathbf{y})}{\min^2_{i}|r_{i,i}|}}$, which corresponds to the complexity upper bounded by $|S|< n\cdot e^{\frac{2\pi d^2(\mathbf{\Lambda},\mathbf{y})}{\min^2_{i}|r_{i,i}|}}$.
\end{my2}

From Theorems 3 \& 4, a tractable decoding trade-off of equivalent SD (i.e., Fincke-Pohst SD) is obtained. Thanks to the usages of $K$ and $\sigma$ from lattice Gaussian distribution, an insightful way is provided to reexamine SD, where the classic
Fincke-Pohst SD or equivalent SD is only a reduced version of the SD based on lattice Gaussian distribution without considering the impact of the query point $\mathbf{y}$. More specifically, note that the requirement of equivalent SD shown in (\ref{b12c}) is similar to that of SD based on lattice Gaussian distribution given in (\ref{lattice gaussian distribution1a}) but with $\rho_{\sigma,\mathbf{y}}(\Lambda)=1$.

\section{Regularized Sphere Decoding Based on Klein's Sampling Probability}
Compared to the sphere radius $D_{\text{equivalent}}$, $D_{\text{LGD}}$ is a better choice because the impact of the query point $\mathbf{y}$ is considered as a regularization term. However, since the regularization term $\rho_{\sigma,\mathbf{y}}(\Lambda)$ is difficult to compute or factorize, directly designing the related decoding algorithm for SD based on lattice Gaussian distribution is quite challenging.
For this reason, the regularized SD based on approximated lattice Gaussian distribution is proposed as an alternative to achieve the decoding gain through suboptimal regularization terms.

\subsection{Algorithm Description}
In order to approximate lattice Gaussian distribution $D_{\Lambda,\sigma,\mathbf{y}}(\mathbf{x})$, Klein's sampling probability $P_{\text{Klein}}(\mathbf{x})$ shown in (\ref{klein distribution}) is utilized.
Since $P_{\text{Klein}}(\mathbf{x})$ is the product of 1-dimensional conditional lattice Gaussian distributions, it is easy to compute and factorize, which enables the sub-sampling probabilities $p(\widehat{x}_i)$ for candidate nodes at each decoding layer. Therefore, compared to (\ref{b12c}) of equivalent SD and (\ref{lattice gaussian distribution1a}) of SD based on lattice Gaussian distribution, the proposed regularized SD based on Klein's sampling probability aims to obtain lattice points satisfying
\begin{equation}
P_{\text{Klein}}(\mathbf{x})\hspace{-.3em}=\hspace{-.3em}\frac{e^{-\frac{1}{2\sigma^2}\parallel \mathbf{Rx}-\mathbf{y} \parallel^2}}{\prod^n_{i=1}\hspace{-.2em}\sum_{\widetilde{x}_{n\hspace{-.1em}-\hspace{-.1em}i\hspace{-.1em}+\hspace{-.1em}1}\in\mathbb{Z}}e^{-\frac{1}{2\sigma^2_{n-i+1}}\|x_{n\hspace{-.1em}-\hspace{-.1em}i\hspace{-.1em}+\hspace{-.1em}1}\hspace{-.1em}-\hspace{-.1em}\widetilde{x}_{n\hspace{-.1em}-\hspace{-.1em}i\hspace{-.1em}+\hspace{-.1em}1}\|^2}}\hspace{-.3em}\geq\hspace{-.3em}\frac{1}{K}.
\label{prunethresholdl}
\end{equation}


Specifically, similar to the decoding mechanism of the equivalent SD, given the initial pruning size $K>1$, the pruning size of each candidate node $\widehat{x}^j_i$ follows
\begin{equation}
K(\widehat{x}^j_i)= K(\underline{\widehat{x}^j_{i}})\cdot p(\widehat{x}^j_i),
\label{prunethresholdl}
\end{equation}
where probability $p(\cdot)$ over $\widehat{x}^j_i$ comes from the 1-dimensional conditional Gaussian distribution in (\ref{piq19}).
Based on it, candidate node $\widehat{x}^j_i$ is retained if it satisfies the \emph{pruning threshold}
\begin{equation}
K(\widehat{x}^j_i)\geq 1,
\label{prunethreshold}
\end{equation}
and the pruning decoding proceeds layer by layer in a backwards order until $i=1$.
Accordingly, from (\ref{prunethresholdl}) and (\ref{prunethreshold}), the pruning threshold can also be expressed by
\begin{equation}
p(\widehat{x}^j_i)\geq\frac{1}{K(\underline{\widehat{x}^j_i})}=\frac{1}{K\cdot p(\widehat{x}^j_{i+1})\cdots p(\widehat{x}^j_{n})}.
\label{prunethreshold2}
\end{equation}
Finally, among all the collected decoding candidates, the one with the smallest Euclidean distance is outputted as the decoding solution.

\subsection{Decoding Performance Analysis}
\begin{my2}
Given the initial pruning size $K>1$, lattice points with Klein's sampling probability
\end{my2}
\vspace{-1em}
\begin{equation}
P_{\text{Klein}}(\mathbf{x})\geq \frac{1}{K}
\label{relation41}
\end{equation}
\emph{will be obtained by the proposed regularized SD based on Klein's sampling probability, which corresponds to the sphere radius as}
\begin{equation}
\|\mathbf{Rx}\hspace{-.2em}-\hspace{-.1em}\mathbf{y}\|\hspace{-.2em}\leq\hspace{-.2em} \sigma\hspace{-.1em}\sqrt{2\ln \frac{K}{\prod^n_{i=1}\hspace{-.2em}\rho_{\sigma_{n\hspace{-.1em}-\hspace{-.1em}i\hspace{-.1em}+\hspace{-.1em}1}, \widetilde{x}_{n\hspace{-.1em}-\hspace{-.1em}i\hspace{-.1em}+\hspace{-.1em}1}}(\mathbb{Z})}}\hspace{-.1em}\triangleq\hspace{-.2em} D_{\text{regularized}}.
\label{lattice gaussian distribution1b}
\end{equation}
\begin{proof}
Considering the fact that $P_{\text{Klein}}(\mathbf{x})=\prod^n_{i=1}p(x_{n-i+1})$, in order to achieve (\ref{relation41}), one has to enumerate all the possible nodes satisfying
{\allowdisplaybreaks\begin{flalign}
p(\widehat{x}^j_i)&\geq \frac{1}{K\cdot p(\widehat{x}^j_n)\cdots p(\widehat{x}^j_{i+1})\cdot p(\widehat{x}^j_{i-1})\cdots p(\widehat{x}^j_{1})}\notag\\
&\geq \frac{1}{K\cdot p(\widehat{x}^j_n)\cdots p(\widehat{x}^j_{i+1})}\notag\\
&=\frac{1}{K(\underline{\widehat{x}^j_i})}
\end{flalign}}for $1\leq i\leq n$, which exactly corresponds to the pruning threshold given in (\ref{prunethreshold2}).
Then, by simple derivation, the sphere radius shown in (\ref{lattice gaussian distribution1b}) can be easily got.
\end{proof}

Clearly, instead of $\rho_{\sigma,\mathbf{y}}(\Lambda)$, in the proposed regularized SD the regularization term $\prod^n_{i=1}\rho_{\sigma_{n-i+1}, \widetilde{x}_{n-i+1}}(\mathbb{Z})$ that depends on the specific value of $\widetilde{x}_i$ at each decoding layer is introduced.
Then, based on it, we define the \emph{decoding performance gain} of the regularized SD over the equivalent SD as
\begin{equation}
G\triangleq\frac{D_{\text{regularized}}}{D_{\text{equivalent}}},
\end{equation}
and then we can arrive at the following Theorem.

\begin{my2}
Let $\sigma=\min_{i}|r_{i,i}|/(2\sqrt{\pi})$, the proposed regularized SD achieves a better decoding performance than the equivalent SD with the decoding gain $G$ bounded by
\end{my2}
\vspace{-1em}
\begin{equation}
1\leq G\leq\sqrt{1+\sum^n_{i=1}\left(\frac{|r_{i,i}|}{\min_{i}|r_{i,i}|}\right)\cdot\frac{\pi}{2\ln K}}.
\end{equation}
\begin{proof}
Let us consider the lower bound for the fist. Based on (\ref{mp1}) and (\ref{lattice gaussian distribution1b}), it follows that
{\allowdisplaybreaks\begin{flalign}
G&=\sqrt{1-\frac{\ln \prod^n_{i=1}\rho_{\sigma_{n-i+1}, \widetilde{x}_{n-i+1}}(\mathbb{Z})}{\ln K}}\notag\\
&\overset{(d)}{\geq}\sqrt{1-\frac{\ln \prod^n_{i=1}\rho_{\sigma_{n-i+1}}(\mathbb{Z})}{\ln K}}\notag\\
&\overset{(e)}{\approx}1,
\end{flalign}}where $(d)$ and $(e)$ come from (\ref{v1f}) and (\ref{bound7}) respectively.


On the other hand, as for the upper bound, we have
{\allowdisplaybreaks\begin{flalign}
G&=\sqrt{1-\frac{\ln \prod^n_{i=1}\rho_{\sigma_{n-i+1}, \widetilde{x}_{n-i+1}}(\mathbb{Z})}{\ln K}}\notag\\
&\overset{(f)}{\leq}\sqrt{1-\frac{\ln \left(\prod^n_{i=1}e^{-\frac{d^2(\mathbb{Z}, \widetilde{x}_{n-i+1})}{2\sigma^2_{n-i+1}}}\cdot\rho_{\sigma_{n-i+1}}(\mathbb{Z})\right)}{\ln K}}\notag\\
&\overset{(g)}{\approx}\sqrt{1+\sum^n_{i=1}\left(\frac{d^2(\mathbb{Z}, \widetilde{x}_{n-i+1})}{2\sigma^2}\cdot|r_{n-i+1,n-i+1}|\right)\cdot\frac{1}{\ln K}}\notag\\
&=\sqrt{1+\sum^n_{i=1}\left(d^2(\mathbb{Z}, \widetilde{x}_{n-i+1})\cdot\frac{|r_{n-i+1,n-i+1}|}{\min_{i}|r_{i,i}|}\right)\cdot\frac{2\pi}{\ln K}}\notag\\
&\overset{(h)}{\leq}\sqrt{1+\sum^n_{i=1}\left(\frac{|r_{i,i}|}{\min_{i}|r_{i,i}|}\right)\cdot\frac{\pi}{2\ln K}},
\end{flalign}}where (f) follows the relationship \cite{RegevNP}
\begin{equation}
\rho_{\sigma_i, c_i}(\mathbb{Z})\geq e^{-\frac{d^2(\mathbb{Z}, c_i)}{2\sigma^2_i}}\cdot\rho_{\sigma_i}(\mathbb{Z}),
\label{xpq1c}
\end{equation}
(g) obeys (\ref{bound7}) and (h) holds because of $0\leq d(\mathbb{Z}, \widetilde{x}_i)\leq\frac{1}{2}$.
\end{proof}

\subsection{Decoding Complexity Analysis}
According to the pruning threshold given in (\ref{prunethreshold2}), the searching space of $\widehat{x}_i$ given $\widetilde{x}_i$ can be derived as
\begin{equation}
|\widehat{x}_i-\widetilde{x}_i|\hspace{-.2em}\leq\hspace{-.3em}\sqrt{\hspace{-.2em}2\sigma^2\hspace{-.1em}\ln \hspace{-.2em} \frac{K}{\prod^n_{j=i}\hspace{-.2em}\rho_{\sigma_{n+j-i},\widetilde{x}_{n+j-i}}(\mathbb{Z})}\hspace{-.2em}-\hspace{-.6em}\sum_{j=i\hspace{-.1em}+\hspace{-.1em}1}^n \hspace{-.4em}|r^2_{j,j}|\hspace{-.2em}\cdot\hspace{-.2em} \left| \widehat{x}_j\hspace{-.2em}-\hspace{-.2em}\widetilde{x}_j \right|^2}\hspace{-.2em}/|r_{i,i}|.
\label{pyw1}
\end{equation}
As can be seen clearly, the regularization terms $\rho_{\sigma_i,\widetilde{x}_i}(\mathbb{Z})$'s play an important role in the searching space of $\widehat{x}_i$.
This can be understood in a straightforward way.
With respect to $0\leq d(\mathbb{Z}, \widetilde{x}_i)\leq\frac{1}{2}$, if it is close to $0$, then the searching space of $\widehat{x}_i$ is encouraged to be reduced and vice versa.
Note that different from SD based on lattice Gaussian distribution who takes $\rho_{\sigma,\mathbf{y}}(\Lambda)$ into account at the very beginning, each component of $\prod^n_{i=1}\rho_{\sigma_{n-i+1}, \widetilde{x}_{n-i+1}}(\mathbb{Z})$ is added one by one along the tree traversal level by level.


\begin{my1}
In the proposed regularized SD, for each saved parent candidate node $\underline{\widehat{x}^j_{i}}$ with $K(\underline{\widehat{x}^j_i})\geq1$,
the number of its children candidate nodes at decoding layer $i$ satisfies
\end{my1}
\vspace{-1em}
\begin{equation}
K_{\text{save}}\leq K(\underline{\widehat{x}^j_i})
\label{pj1a}
\end{equation}
\emph{if} $\sigma=\min_{i}|r_{i,i}|/(2\sqrt{\pi})$.
\begin{proof}
We start the proof by considering the cases of $1\leq K(\underline{\widehat{x}^j_i})<2$ and $K(\underline{\widehat{x}^j_i})\geq2$ respectively.

On one hand, when $1\leq K(\underline{\widehat{x}^j_i})<2$, based on the pruning threshold in (\ref{prunethreshold2}), candidate node will be saved if
\begin{equation}
p(\widehat{x}^j_i)\geq\frac{1}{K(\underline{\widehat{x}^j_i})}>\frac{1}{2}.
\end{equation}
Clearly, because $p(\cdot)$ is a distribution with $\sum_jp(\widehat{x}^j_i)=1$, there is at most one integer candidate satisfying above requirement, which implying
\begin{equation}
K_{\text{save}}\leq1\leq K(\underline{\widehat{x}^j_i})
\end{equation}
no matter what $\sigma>0$ is.

On the other hand, when $K(\underline{\widehat{x}^j_i})\geq2$, according to the pruning threshold in (\ref{prunethreshold2}), the condition shown in (\ref{pj1a}) holds if and only if the $\lfloor K(\underline{\widehat{x}^j_i})+1\rfloor$th closest integer candidate to $\widetilde{x}_i$ will definitely be pruned, that is
\begin{equation}
K(\underline{\widehat{x}^j_i})p(\widehat{x}^{\lfloor K(\underline{\widehat{x}^j_i})+1 \rfloor}_i)< 1.
\label{wo1a1}
\end{equation}

Then, from (\ref{wo31}), (\ref{wo1a1}) can be achieved if
\begin{equation}
K(\underline{\widehat{x}^j_i})\cdot e^{-\frac{1}{8\sigma_i^2}(\lfloor K(\underline{\widehat{x}^j_i})+1\rfloor-1)^2}<\rho_{\sigma_i,\widetilde{x}_i}(\mathbb{Z}).
\label{wo4a}
\end{equation}

Moreover, according to the relationship shown in (\ref{xpq1c}), (\ref{wo4a}) holds if
\begin{equation}
K(\underline{\widehat{x}^j_i})\cdot e^{-\frac{1}{8\sigma_i^2}(\lfloor K(\underline{\widehat{x}^j_i})+1\rfloor-1)^2}<e^{-\frac{d^2(\mathbb{Z}, \widetilde{x}_i)}{2\sigma^2_i}}\cdot\rho_{\sigma_i}(\mathbb{Z})
\label{wo4b}
\end{equation}
is fulfilled. Because of $0\leq d(\mathbb{Z}, \widetilde{x}_i)\leq1/2$, (\ref{wo4b}) can be further derived as
\begin{equation}
\sigma^2<\frac{(\lfloor K(\underline{\widehat{x}^j_i})+1\rfloor-1)^2-1}{8\ln \frac{K(\underline{\widehat{x}^j_i})}{\rho_{\sigma_i}(\mathbb{Z})}}\cdot\|\widehat{\mathbf{b}}_i\|^2.
\label{wo5a}
\end{equation}


Finally, it is straightforward to verify that (\ref{wo5a}) is satisfied when $\sigma=\min_{i}|r_{i,i}|/(2\sqrt{\pi})$ (i.e., $\rho_{\sigma_i}(\mathbb{Z})\approx1$), completing the proof.
\end{proof}

\begin{my2}
In the proposed regularized SD, let $\sigma=\min_{i}|r_{i,i}|/(2\sqrt{\pi})$, the number of visited nodes is upper bounded by
\end{my2}
\vspace{-2em}
\begin{equation}
|S|< nK,
\label{hp6a}
\end{equation}
\emph{and the number of collected decoding candidates at the end is upper bounded by}
\begin{equation}
|L|< K.
\label{hp6c}
\end{equation}
\begin{proof}
First of all, according to Lemma 4, the number of saved candidate nodes at each layer is upper bounded by the summation of pruning sizes at the previous layer, namely,
\begin{equation}
K^{\text{layer}\ i}_{\text{save}}=\sum K_{\text{save}}\leq \sum K(\underline{\widehat{x}^j_i})=K^{\text{layer}\ i+1}_{\text{pruning\ size}}.
\end{equation}

Meanwhile, according to (\ref{prunethresholdl}), we have
\begin{equation}
\sum_jK(\widehat{x}^j_i)=K(\underline{\widehat{x}^j_i})\cdot\sum_jp(\widehat{x}^j_i)<K(\underline{\widehat{x}^j_i}),
\end{equation}
which means the summation of pruning sizes at each decoding layer is decreasing from layer $n$ to 1, i.e.,
\begin{equation}
K^{\text{layer}\ 1}_{\text{pruning\ size}}<\ldots< K^{\text{layer}\ n}_{\text{pruning\ size}}< K^{\text{layer}\ n+1}_{\text{pruning\ size}}=K.
\label{pj1pa1}
\end{equation}

Therefore, the number of visited nodes in the proposed regularized SD is upper bounded by
\begin{equation}
|S|=\sum_iK^{\text{layer}\ i}_{\text{save}}\leq \sum_iK^{\text{layer}\ i+1}_{\text{pruning\ size}}< nK.
\end{equation}

On the other hand, since the number of collected decoding candidates $|L|$ corresponds to $K^{\text{layer}\ 1}_{\text{save}}$, it is upper bounded by
\begin{equation}
|L|< K,
\end{equation}
completing the proof.
\end{proof}

Based on Theorems 6 \& 7, given the same complexity upper bound, the regularized SD has a better decoding performance than the equivalent SD (i.e., Fincke-Pohst SD) for arbitrary query point $\mathbf{y}$, thus leading to a better decoding trade-off.
\begin{my2}
The initial pruning size $K$ of solving the CVP by the proposed regularized SD is upper bounded by
\end{my2}
\vspace{-1em}
\begin{equation}
K\leq e^{\frac{2\pi d^2(\mathbf{\Lambda},\mathbf{y})}{\min^2_{i}|r_{i,i}|}},
\end{equation}
\emph{which corresponds to the complexity upper bounded by }
\begin{equation}
|S|< n\cdot e^{\frac{2\pi d^2(\mathbf{\Lambda},\mathbf{y})}{\min^2_{i}|r_{i,i}|}}.
\end{equation}
\begin{proof}
Given the Euclidean distance $d(\mathbf{\Lambda},\mathbf{y})$ between the lattice $\Lambda$ and the query point $\mathbf{y}$, then CVP will be solved by setting $D_{\text{regularized}}=d(\mathbf{\Lambda},\mathbf{y})$, which corresponds to
\begin{equation}
\sigma\sqrt{2\ln \frac{K}{\prod^n_{i=1}\rho_{\sigma_{n-i+1},\widetilde{x}_{n-i+1}}(\mathbb{Z})}}= d(\mathbf{\Lambda},\mathbf{y}).
\end{equation}
Moreover, by letting $\sigma=\min_{i}|r_{i,i}|/(2\sqrt{\pi})$,  it follows that
{\allowdisplaybreaks\begin{flalign}
K&=\left(\prod^n_{i=1}\rho_{\sigma_{n-i+1},\widetilde{x}_{n-i+1}}(\mathbb{Z})\right)\cdot e^{\frac{2\pi d^2(\mathbf{\Lambda},\mathbf{y})}{\min^2_i|r_{i,i}|}}\notag\\
&\leq\left(\prod^n_{i=1}\rho_{\sigma_{n-i+1}}(\mathbb{Z})\right)\cdot e^{\frac{2\pi d^2(\mathbf{\Lambda},\mathbf{y})}{\min^2_i|r_{i,i}|}}\label{bvw1}\\
&\approx e^{\frac{2\pi d^2(\mathbf{\Lambda},\mathbf{y})}{\min^2_i|r_{i,i}|}}.
\end{flalign}}On the other hand, according to (\ref{hp6a}) given in Theorem 7, the complexity is upper bounded by
\begin{equation}
|S|< n\cdot e^{\frac{2\pi d^2(\mathbf{\Lambda},\mathbf{y})}{\min^2_i|r_{i,i}|}},
\end{equation}
completing the proof.
\end{proof}

Here, we point out that the upper bound shown above is rather loose due to the inequality (\ref{v1f}) applied in (\ref{bvw1}).
Additionally, it should be noticed that the ML decoding criterion in (\ref{eqn:ML DecodingQR1}) motivated by lattice Gaussian distribution could reduce to the one in (\ref{eqn:ML DecodingQR}). This happens when $\sigma$ is sufficiently large, the impact of the parameter $\sigma$ actually turns out to be removed in a uniform distribution. For this reason, it is not encouraged to increase $\sigma$ intensively for solving the CVP so that the benefit from $\sigma$ by enlarging the difference of Euclidean distance $\|\mathbf{R\widehat{x}}-\mathbf{y}\|$ can be well explored.
\begin{my7}
The ML decoding criterion $\widehat{\mathbf{x}}_{\text{ML}}=\underset{\mathbf{x}\in \mathbb{Z}^{n}}{\operatorname{arg~max}} \, D_{\Lambda,\sigma,\mathbf{y}}(\mathbf{x})$ will reduce to $\widehat{\mathbf{x}}_{\text{ML}}=\underset{\mathbf{x}\in \mathbb{Z}^{n}}{\operatorname{arg~min}} \, \|\mathbf{R}\mathbf{x}-\mathbf{y}\|^2$ with $\sigma\rightarrow\infty$.
\end{my7}

\section{Further Enhancement of the Regularized SD}
Because the summation of the pruning sizes is decreasing layer by layer, both the equivalent SD and the regularized SD have a same problem: they only work well when the initial pruning size $K$ is large enough.
Given a small size $K$, the pruning decoding still works but it may terminate at early decoding layers as all the possible candidate nodes are discarded due to limited pruning size $K(\widehat{x}^j_i)$.
This is similar to Fincke-Pohst SD, where no decoding solution will output by a small sphere radius $D$.
In this case, although considerable complexity cost has been consumed in the early decoding layers, no lattice points will be reversed, rendering the decoding meaningless.
This actually raises a natural question: how to fully exploit the decoding potential with a small $K$?
Therefore, in what follows, we try to answer this question via another decoding criterion designed for candidate saving, and we refer to it as \emph{candidate protection}.
Note that the proposed candidate protection is also applicable to equivalent SD but here we only discuss it through the regularized SD.

\subsection{Candidate Protection}
In essence, as for candidate nodes with small pruning size $K(\widehat{x}^j_i)$, the usage of candidate protection tries to rescue the most valuable decoding candidate along that branch, and
the decoding solution consists of the closest candidate nodes $\widehat{x}^1_{i's}$ in the rest of layers normally has the largest sampling probability in statistics.
Therefore, candidate protection effectively extends the pruning threshold in (\ref{prunethreshold}) by returning some valuable decoding candidates rather than removing them directly.
Specifically, with respect to candidate node $\widehat{x}^j_i$ with small pruning size
\begin{equation}
2> K(\widehat{x}^j_i)\geq 1,
\end{equation}
candidate protection is activated to obtain the closest candidate nodes $\widehat{x}^1_{i-1}, \ldots, \widehat{x}^1_1$ in the rest of decoding layers, which directly yields a candidate lattice point $\widehat{\mathbf{x}}$:
\begin{equation}
\widehat{\mathbf{x}}=[\ \overbrace{\underbrace{\widehat{x}^1_1\ ,\ \ldots ,\ \widehat{x}^1_{i-1}}_{\text{candidate protection}}\ , \hspace{-.8em}\underbrace{\widehat{x}^{j}_i}_{2> K(\cdot)\geq 1}\hspace{-1em},\ \underbrace{\widehat{x}^{j}_{i+1},\ \ldots,\ \widehat{x}^{j}_n}_{K(\cdot)\geq2}}^{\longleftarrow\text{decoding order}}\ ]^T.
\label{pwu1}
\end{equation}

We point out that the pruning threshold is smoothly compatible with candidate protection as the latter tries to activate a few candidate nodes discarded by the former.
Intuitively, the proposed criterion of candidate protection extends the initial pruning size $K>1$ to $K\geq1$, and it is easy to verify that the decoding performance of Babai's nearest plane algorithm will be achieved for the case of $K=1$.
Meanwhile, candidate protection works with respect to the candidate nodes from decoding layer $n$ to $2$ as the saved nodes at decoding layer $i=1$ will be outputted rather than discarded.
Moreover, candidate protection can be simply carried out through Babai's nearest plane algorithm since $\widehat{x}^1_1,...,\widehat{x}^1_n$ is just the decoding result of it.
Counterintuitively, even for a large enough $K$, decoding candidates coming from candidate protection also have the contribution in outputting the optimal decoding solution.
This is due to the application of Klein's sampling probability, where distortion does exist compared to the exact lattice Gaussian distribution.

To summarize, at each decoding layer, the proposed regularized SD based on Klein's sampling probability algorithm operates in the following two steps:
\begin{itemize}
  \item \emph{Calculate the pruning size $K(\widehat{x}_i^j)$ by (\ref{prunethresholdl})}.
  \item \emph{Obtain candidate nodes $\widehat{x}_i^j$ by (\ref{prunethreshold}). If $2> K(\widehat{x}^j_i)\geq 1$, invoke Babai's nearest plane algorithm to directly return a decoding candidate $\widehat{\mathbf{x}}$}.
\end{itemize}
For a better understanding, an illustration of the proposed regularized SD algorithm is presented in Fig. 2 with more details.

\begin{figure*}[t]
\includegraphics[width=7in,height=3.3in]{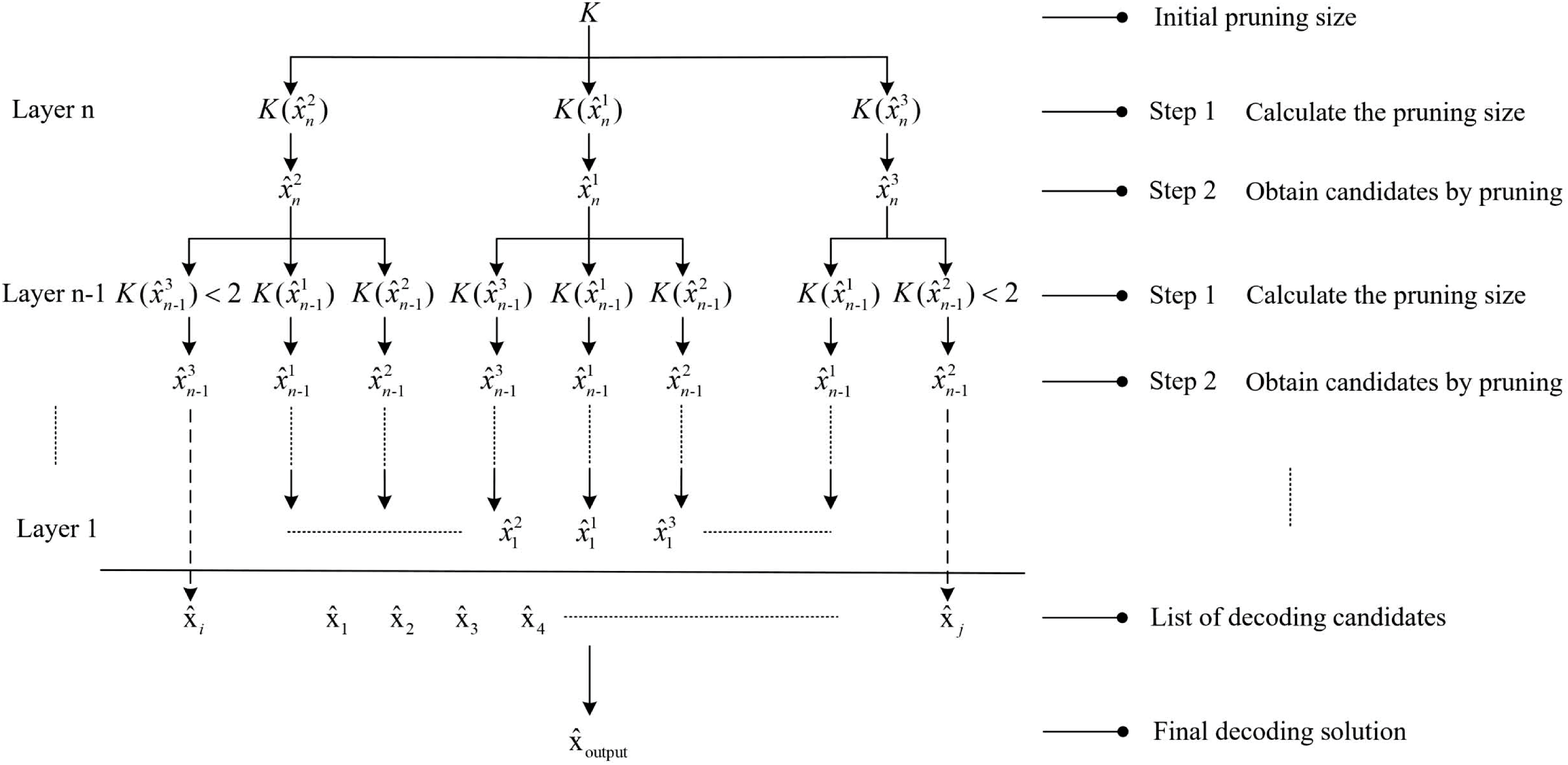}
\caption{Illustration of the proposed regularized sphere decoding algorithm, where $K(\widehat{x}^j_i)\geq1$ and the dashed line stemmed from $K(\widehat{x}^j_i)<2$ denotes the closest candidate nodes $\widehat{x}^1_{i-1}, \ldots, \widehat{x}^1_1$ in the rest of layers are retained to directly yield a decoding candidate $\widehat{\mathbf{x}}$.}
\label{IDD figure}
\end{figure*}

\subsection{Decoding Complexity Analysis}
On the other hand, interestingly, even with the application of candidate protection, we show that the complexity $|S|$ as well as number of collected decoding candidates $|L|$ in the proposed regularized SD still maintains the same upper bound as before.
\begin{my2}
Given the initial pruning size $K\geq1$, the number of decoding candidates outputted by the proposed regularized SD algorithm (considering both pruning threshold and candidate protection) is upper bounded by
\end{my2}
\vspace{-1em}
\begin{equation}
|L|< K
\label{pj1pala}
\end{equation}
\emph{with bounded total number of visited nodes}
\begin{equation}
|S|< nK
\label{pj1palb}
\end{equation}
\vspace{-.5em}
\emph{for} $\sigma=\min_{i}|r_{i,i}|/(2\sqrt{\pi})$.
\vspace{1em}
\begin{proof}
The collected decoding lattice points come from pruning threshold and candidate protection respectively. As shown in (\ref{pj1pa}), the summation of pruning size at each layer is decreasing, which can be further expressed as
{\allowdisplaybreaks\begin{flalign}
K&=K(\underline{x^j_n}) \notag\\
&>\sum K(x^{\text{terminate}}_n)+\sum K(x^{\text{pruning}}_n) \notag\\
&>\sum K(x^{\text{terminate}}_n)+\sum K(x^{\text{terminate}}_{n-1})+\sum K(x^{\text{pruning}}_{n-1}) \notag\\
&>\cdots\notag\\
&>\sum_{i=2}^n\left[\sum K(x^{\text{terminate}}_i)\right]+\sum K(x^{\text{pruning}}_{2}),
\end{flalign}}with $2>K(x^{\text{terminate}}_i)\geq1$ and $K(x^{\text{pruning}}_{2})\geq2$.

\floatname{algorithm}{Algorithm}

\setcounter{algorithm}{2}

\begin{algorithm}[t]
\caption{Regularized SD Algorithm}
\begin{algorithmic}[1]
\Require
$K, \mathbf{R}, \mathbf{y}, \sigma=\min_{i}|r_{i,i}|/(2\sqrt{\pi}), L=\emptyset$
\Ensure
$\mathbf{Rx}\in\Lambda$
\State invoke \textbf{Function 1} with $i=n$ to decode layer by layer
\State add all the candidates $\widehat{\mathbf{x}}$'s generated by \textbf{Function 2} to $L$
\State output $\widehat{\mathbf{x}}=\underset{\mathbf{x}\in L}{\operatorname{arg~min}} \, \|\mathbf{y}-\mathbf{R}\mathbf{x}\|$ as the decoding solution
\end{algorithmic}
\end{algorithm}

\floatname{algorithm}{Function}
%
\setcounter{algorithm}{1}

\begin{algorithm}[t]
\caption{Pruning Decoding at layer $i$ given $[\widehat{x}_n,\ldots,\widehat{x}_{i+1}]$}
\begin{algorithmic}[1]
\State compute $\widetilde{x}_i$ according to (\ref{eqn:noise effection in determination})
\State compute probability $p(\widehat{x}^j_i)$ by (\ref{piq19}) with $j\in[1,2,3]$
\State compute pruning size $K(\widehat{x}^j_i)$ according to (\ref{prunethresholdl})
\For {each specific integer candidate $\widehat{x}^j_i$}
\If {$K(\widehat{x}^j_i)<1$}
\State \hspace{-2em}prune $\widehat{x}^j_i$ from the tree-search decoding
\Else
\State {\hspace{-2em}save $\widehat{x}^j_i$} to form the decoding result $[\widehat{x}_n,\ldots,\widehat{x}_{i+1}, \widehat{x}^j_i]$
\If {$2> K(\widehat{x}^j_i)\geq 1$}
\State \hspace{-2.5em}decode the rest of layers by SIC to get a candidate $\widehat{\mathbf{x}}$
\ElsIf {$K(\widehat{x}^j_i)\geq2$}
\If {$i=1$}
\State \hspace{-2em}output the candidate $\widehat{\mathbf{x}}$
\Else
\State \hspace{-2em}invoke \textbf{Function 2} to decode the next layer $i-1$
\EndIf
\EndIf
\EndIf
\EndFor
\end{algorithmic}
\end{algorithm}

Based on the configuration of candidate protection, only one decoding candidate will be saved for each $K(x^{\text{terminate}}_i)$, $2\leq i\leq n$, which means the number of collected decoding candidates generated by candidate protection from decoding layer $n$ to $2$ is bounded by
\begin{equation}
|L_{\text{terminate}}|\leq\sum_{i=2}^n\left[\sum K(x^{\text{terminate}}_i)\right].
\label{k1a}
\end{equation}

Besides, the number of decoding candidates collected by the pruning threshold corresponds to $K^{\text{layer}\ 1}_{\text{save}}$, which is upper bounded by
\begin{equation}
|L_{\text{pruning}}|=K^{\text{layer}\ 1}_{\text{save}}\leq \sum K(x^{\text{pruning}}_{2})
\label{k1b}
\end{equation}
according to (\ref{pj1a}) in Lemma 4.
Therefore, based on (\ref{k1a}) and (\ref{k1b}), it follows that
\begin{equation}
|L|=|L_{\text{pruning}}|+|L_{\text{terminate}}|<K.
\end{equation}
Consequently, as all the visited nodes are taken into account to generate $|L|$ decoding candidates, the number of visited candidate nodes can be easily derived as
\begin{equation}
|S|< nK,
\end{equation}
completing the proof.
\end{proof}

\begin{figure*}[t]
\begin{center}
\includegraphics[width=6.5in]{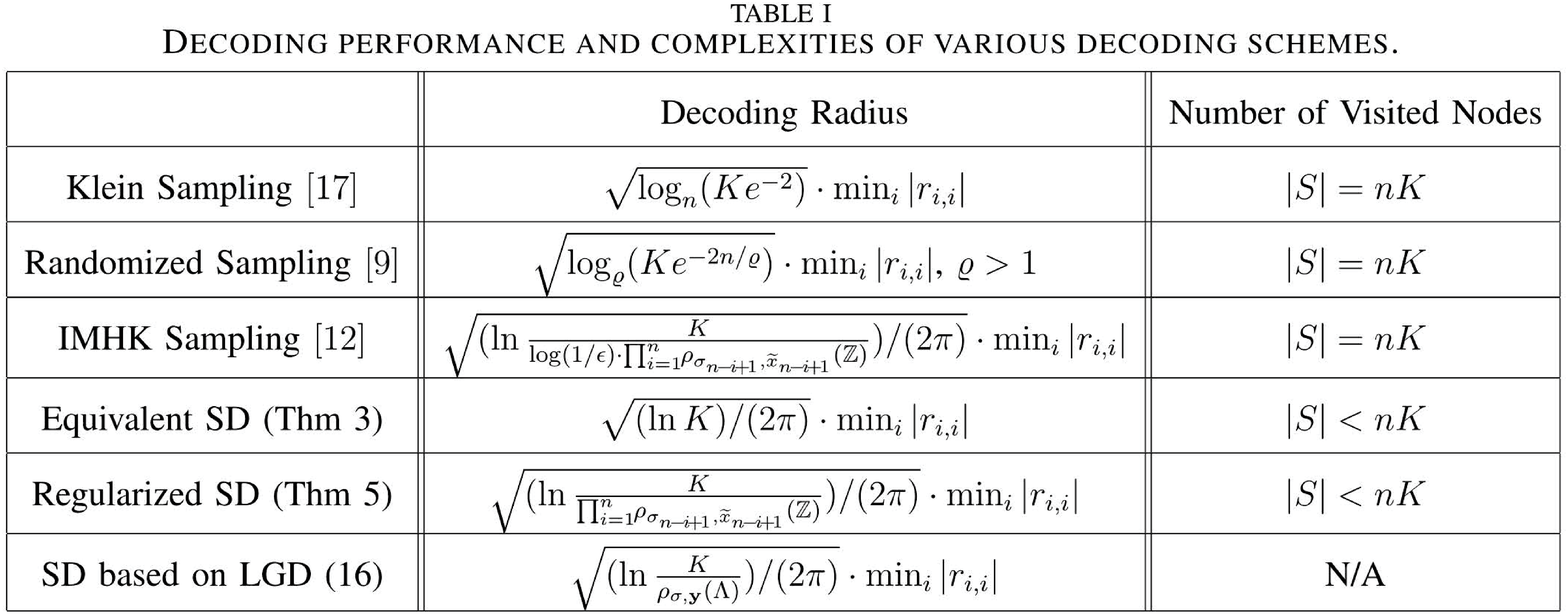}
\end{center}
\end{figure*}

Here, for a better understanding, a comparison about the average number of decoding candidates $|L|$ obtained by the proposed regularized SD algorithm with $\sigma=\min_{i}|r_{i,i}|/(2\sqrt{\pi})$ for various uncoded MIMO systems using 64-QAM is given in Fig. 3. Clearly, $|L|$ is always much smaller than the initial pruning size $K$ in all cases. Specifically, with the increment of $K$, $|L|$ improves gradually as more qualified lattice points are obtained by the relaxed pruning threshold. On the other hand, with the improvement of system dimension, the tree-structure becomes deeper while the decoding potential of the regularized SD algorithm will be further exploited, which leads to a larger $|L|$ in average.
To summarize, decoding radii and numbers of visited nodes of various decoding schemes are given in Table 1. As it can be seen, regularized SD not only achieves a better decoding trade-off than equivalent SD, but also outperforms sampling decoding schemes in both decoding performance and complexity.
Here we emphasize again that the above decoding radii of equivalent SD and regularized SD algorithms are only based on the decoding candidates collected by criterion 1, which means the real decoding performance from both criterions 1 and 2 could be better.

The proposed regularized SD is closely related with randomized sampling given in \cite{CongDecoding}. In fact, it can be viewed as a deterministic version of the randomized sampling decoding, which performs the sampling from Klein's sampling probability individually.
Note that $K$ in sampling algorithms denotes the number of individual sampling operations followed by $K$ decoding candidates while $K$ in the proposed statistical lattice pruning decoding algorithm only serves as a nominal pruning size with the number of decoding candidates $|L|<K$, which implies the great potential of complexity reduction.
Different from sampling decoding algorithms, there is no repetition in the final $|L|$ decoding candidates of the regularized SD. Meanwhile, thanks to the tree-search structure, computational complexity for each candidate node is performed only once, thus saving a lot of complexity. Moveover, due to the deterministic way of collecting lattice points, the performance loss suffered from randomness during the sampling can be avoided. Therefore, compared to sampling decoding algorithms, a better decoding trade-off is achieved by the regularized SD with both performance improvement and complexity reduction.
On the other hand, the computational complexity of every single sampling in randomized sampling decoding is $O(n^2)$ (corresponds to $n$ visited nodes) and the total computational complexity by returning $K$ samples is $O(K\cdot n^2)$ (corresponds to $nK$ visited nodes). This actually can be viewed as an upper bound of the computational complexity of the regularized SD with less than $nK$ visited nodes.

\begin{my7}
Given the initial pruning size $K\geq1$ and standard deviation $\sigma=\min_{i}|r_{i,i}|/(2\sqrt{\pi})$, the computational complexity of the regularized SD is $O(K\cdot n^2)$.
\end{my7}

\begin{figure}[t]
\includegraphics[width=3.5in]{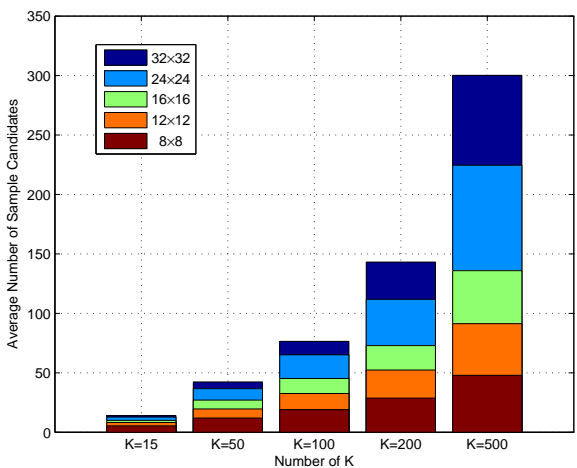}
\vspace{-1em}
  \caption{Average number of decoding candidates $|L|$ versus number of $K$ for various uncoded MIMO systems using 64-QAM at SNR per bit = 17dB.}
  \label{simulation 4}
\end{figure}


\subsection{Further Complexity Reduction in Implementation}
We now investigate the size of index $j$ in the regularized SD.
In fact, from (\ref{piq19}), the sampling probability of $j$th candidate node at decoding layer $i$ can be written as
\begin{equation}
p(x^j_i)\hspace{-.2em}=\hspace{-.2em}\begin{cases}e^{-\frac{1}{2\sigma_i^2}((j-1)/2+d)^2}\hspace{-.5em}/\rho_{\sigma_i, \widetilde{x}_i}(\mathbb{Z})\ \text{when}\ j\ \text{is odd}, \\
       e^{-\frac{1}{2\sigma_i^2}(\frac{j}{2}-d)^2}\hspace{-.5em}/\rho_{\sigma_i, \widetilde{x}_i}(\mathbb{Z})\ \ \ \ \ \ \ \text{when}\ j\ \text{is even},
       \end{cases}
\label{pwq1}
\end{equation}
where $\frac{1}{2}\geq d=|x^1_i-\widetilde{x}_i|\geq0$.
Therefore, the summation probability of the first $2N$ candidate nodes with respect to $\widetilde{x}_i$ can be expressed as
\begin{equation}
P_{2N}\hspace{-.2em}=\hspace{-.2em}\sum^N_{j=1}\hspace{-.2em}\left(e^{-\frac{1}{2\sigma_i^2}(j-1+d)^2}\hspace{-.5em}+e^{-\frac{1}{2\sigma_i^2}(j-d)^2}\right)\hspace{-.2em}/\rho_{\sigma_i, \widetilde{x}_i}(\mathbb{Z}).
\end{equation}


Then, with $\sigma=\min_{i}|r_{i,i}|/(2\sqrt{\pi})$, the probability except those $2N$ candidate nodes can be derived as
{\allowdisplaybreaks\begin{flalign}
1-P_{2N}&=\sum_{j\geq N+1}\hspace{-.4em}\left(e^{-\frac{1}{2\sigma_i^2}(j-1+d)^2}+e^{-\frac{1}{2\sigma_i^2}(j-d)^2}\right)\hspace{-.2em}/\rho_{\sigma_i, \widetilde{x}_i}(\mathbb{Z}) \notag\\
&<\sum_{j\geq N+1}\hspace{-.4em}2\cdot e^{-\frac{1}{2\sigma_i^2}(j-1)^2}\hspace{-.4em}/\rho_{\sigma_i, \widetilde{x}_i}(\mathbb{Z}) \notag\\
&<\sum_{j\geq N+1}2\cdot e^{-\frac{1}{2\sigma_i^2}[(j-1)^2-\frac{1}{4}]} /\rho_{\sigma_i}(\mathbb{Z})\notag\\
&\approx\sum_{j\geq N+1}2\cdot e^{-2\pi[(j-1)^2-\frac{1}{4}]} \notag\\
&=O\left(e^{-2\pi N^2}\right),
\label{xpq1}
\end{flalign}}which implies the tail bound (\ref{xpq1}) decays exponentially fast due to $e^{2\pi}\gg1$.

\begin{my5}
From (\ref{xpq1}), with $\sigma=\min_{i}|r_{i,i}|/(2\sqrt{\pi})$, the number of children candidate nodes is severely limited due to the negligible probabilities $p(x^j_i), j>3$.
\end{my5}

Therefore, in practice, $j=3$ is recommended since probability computation of  $p(x^j_i), j>3$ is meaningless unless the initial pruning size $K$ is sufficiently large.

\section{Simulation}
In this section, the performance and complexity of the proposed sphere decoding-based algorithms are evaluated in the large-scale MIMO detection.
Specifically, the $i$th entry of the transmitted signal $\mathbf{x}$, denoted as $x_i$, is a modulation symbol taken independently from an $M$-QAM constellation $\mathcal{X}$
with Gray mapping. Meanwhile, we assume a flat fading environment, where the square channel matrix
$\mathbf{H}$ contains uncorrelated complex Gaussian fading gains with unit
variance and remains constant over each frame duration. Let $E_b$ represents the average power per bit at the receiver, then the signal-to-noise ratio (SNR) $E_b/N_0=n/(\text{log}_2(M)\sigma_w^2)$ where $M$ is the modulation level and $\sigma_w^2$ is the noise variance. Then, we can express the system model as
\begin{equation}
\mathbf{c}=\mathbf{H}\mathbf{x}+\mathbf{w}.
\label{eqn:System Model3}
\end{equation}

\begin{figure}[t]
\includegraphics[width=3.5in]{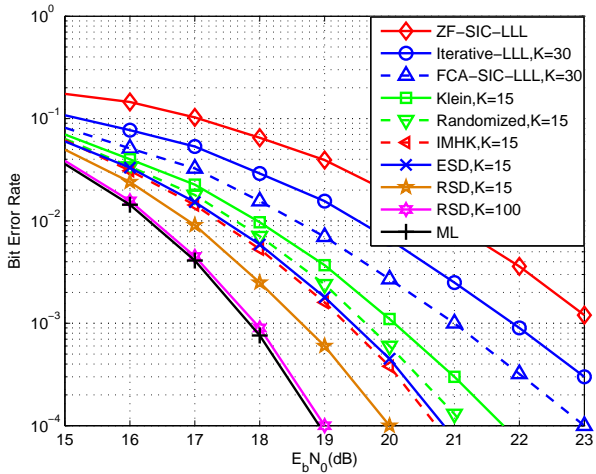}
\vspace{-1em}
  \caption{Bit error rate versus average SNR per bit for the uncoded $12 \times 12$ MIMO system using 64-QAM.}
  \label{simulation 2a}
\end{figure}


\begin{figure}[t]
\includegraphics[width=3.5in]{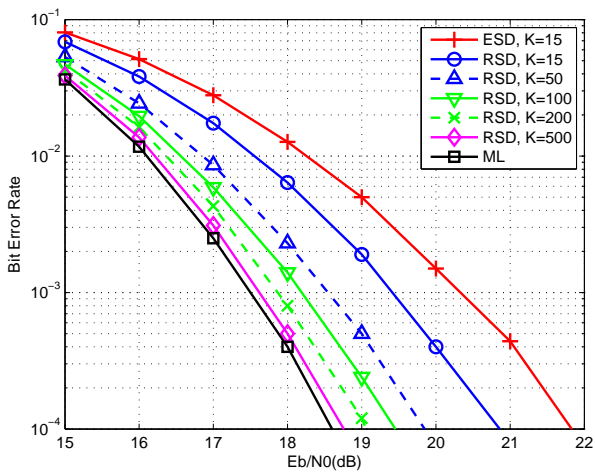}
\vspace{-1em}
  \caption{Bit error rate versus average SNR per bit for the uncoded $16 \times 16$ MIMO system using 64-QAM.}
  \label{simulation 2b}
\end{figure}

Fig. \ref{simulation 2a} shows the bit error rate (BER) of the proposed equivalent sphere decoding (ESD) and regularized sphere decoding (RSD) algorithms compared with other decoding schemes in a $12\times12$ uncoded MIMO system with 64-QAM. Here, lattice-reduction-aided SIC decoding serves as a performance baseline and the ML decoding is implemented by the Schnorr-Euchner strategy from \cite{DamenDetectionSearch}. Clearly, compared to the fixed candidates algorithm (FCA) in \cite{MaxiaoliSoft} and iterative list decoding in \cite{Shimokawa} with 30 samples, sampling decoding algorithms such as Klein's sampling decoding \cite{Klein}, randomized sampling decoding \cite{CongRandom} and IMHK sampling decoding \cite{ZhengWangTIT17} offer not only the improved BER performance but also the promise of smaller sample size $K$.
Besides the help of LLL, all the sampling decoding schemes here including the proposed ESD and RSD are also enhanced by the MMSE augmentation strategy, where more details can be found in \cite{CongRandom}.
As expected, RSD achieves a better BER performance than those sampling decoding schemes with the same $K$. With the increment of $K$, the BER performance improves gradually. Meanwhile, there is a substantial performance gap between RSD and ESD. Observe that with $K=100$, the performance of the RSD algorithm suffers negligible loss compared with ML. Therefore, with a moderate $K$, near-ML performance can be achieved.


In order to show the performance comparison with different initial pruning size $K$, Fig. \ref{simulation 2b} is given to illustrate the BER performance of ESD and RSD in a $16\times16$ uncoded system with 64-QAM. According to (\ref{relation41}) in Theorem 5, a larger $K$ leads to a larger sphere radius $D_{\text{regularized}}$, which corresponds to a better decoding performance. More specifically, as shown in (\ref{relation41}), a larger $K$ naturally corresponds to a looser pruning threshold, which allows more lattice points to be obtained. Therefore, as can be seen clearly, with the increment of $K$, the BER performance improves gradually to the ML decoding performance. It is interesting that in Fig. \ref{simulation 2a} near-ML decoding performance can be achieved with $K=100$ while in Fig. \ref{simulation 2b} near-ML decoding performance requires $K=500$. This is because the larger system dimension has a deeper tree-structure to decode, which requires more initial pruning size $K$ to explore. Note that the number of candidate lattice points $|L|$ is upper bounded by $K$, and the complexity increment with respect to $K$ is mild as expected, thus resulting in a promising trade-off between performance and complexity.


\begin{figure}[t]
\includegraphics[width=3.5in]{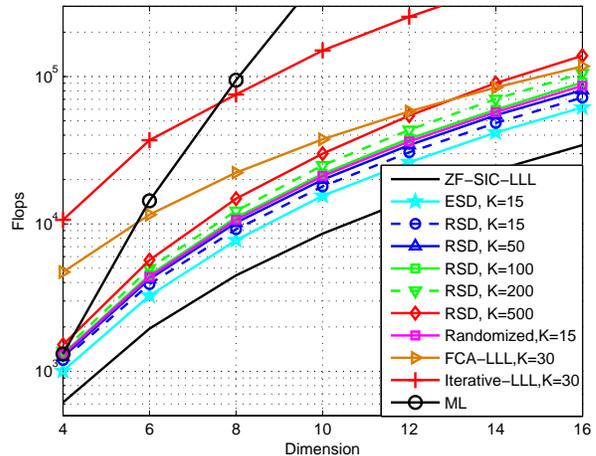}
\vspace{-1em}
  \caption{Complexity comparison in flops for the uncoded MIMO system using 64-QAM at SNR per bit = 17dB.}
  \label{simulation 2}
\end{figure}

\begin{figure}[t]
\includegraphics[width=3.5in]{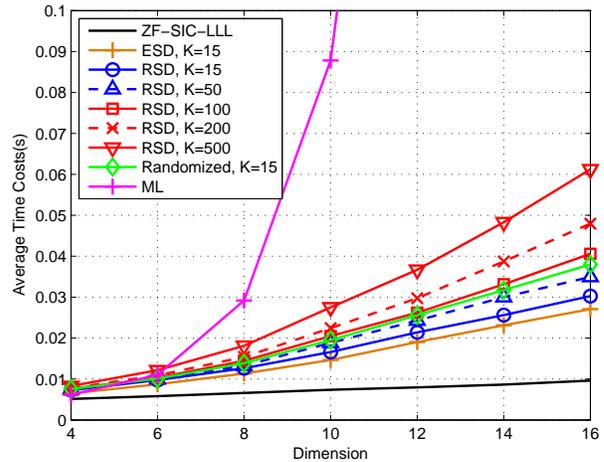}
\vspace{-1em}
  \caption{Complexity comparison in average time cost for the uncoded MIMO system using 64-QAM at SNR per bit = 17dB.}
  \label{simulation 3}
\end{figure}

Fig. \ref{simulation 2} shows the complexity comparison in flops of the proposed ESD and RSD algorithm with other decoding schemes in different system dimensions, where the flops evaluation scenario that we use comes from \cite{MatlabFlops}. Clearly, in the uncoded MIMO system with 64-QAM, the proposed ESD and RSD need much lower flops than other decoding schemes under the same size $K$. This benefit comes from the adaptation of the tree-structure from conventional pruning algorithm, which reduces the computation in sampling procedures by removing all the unnecessary repetitions and calculations.
Specifically, the flops cost of RSD with $K=50$ is less than that of randomized sampling decoding with $K=15$.
More importantly, with the increase of $K$, the decoding performance improves gradually but the complexity increment is rather mild, which is in line with Corollary 4.
Consequently, better BER performance and less complexity requirement make the RSD algorithm very promising for solving the CVP.


Following the same scenario in Fig. \ref{simulation 2}, as a complement to illustrate the computational cost, Fig. \ref{simulation 3} is given to show the complexity comparison in average elapsed running times.
In particular, the uncoded MIMO system takes 64-QAM at SNR per bit = 17dB, and the simulation is conducted by MATLAB R2016a on a single computer, with an Intel Core i7 processor at 2.7GHz, a RAM of 8GB and Windows 10 Enterprise Service Pack operating system.
As can be seen clearly, the average elapsed running time of SIC-LLL decoding scheme increases slightly with the increase of system dimension. On the contrary, the optimal ML decoding from \cite{DamenDetectionSearch} schemes takes an exponentially increasing average elapsed running time, which is unaffordable in most of cases. As expected, under the same $K$, the proposed RSD has a lower average elapsed running time than randomized sampling decoding, making it easy to be implemented especially in high-dimensional MIMO systems.


\section{Conclusions}

In this paper, lattice Gaussian distribution was introduced into sphere decoding for solving the CVP in lattice decoding. Different from the conventional SD, the sphere radius of the proposed SD based on lattice Gaussian distribution is characterized by the initial pruning size $K$, the standard deviation $\sigma$ and the Gaussian scalar $\rho_{\sigma,\mathbf{y}}(\Lambda)$, which provides more freedom to examine the mechanism of SD. Based on it, we showed that the proposed equivalent SD is exactly the same with the classic Fincke-Pohst SD but with a tractable decoding trade-off between performance and complexity. Moreover, to further exploit the decoding potential, the regularized SD was proposed to fully take advantages of these regularization terms. By approximating the lattice Gaussian distribution with Klein's sampling probability, a larger size of sphere radius can be achieved under the same complexity constraint, leading to a better decoding trade-off than equivalent SD. In addition, another decoding criterion referred to as candidate guard was given as well, which generalizes the proposed SD algorithms from ML decoding to BDD. By doing this, the proposed SD algorithms suit well for the various decoding requirements, where the decoding trade-off is adjusted through the single parameter $K$ freely.

\appendices

\bibliographystyle{IEEEtran}
\bibliography{IEEEabrv,reference1}

\end{document}